\documentclass[fleqn,usenatbib]{mnras}
\usepackage[pdfpagelabels=false]{hyperref}	
\hypersetup{colorlinks=true,linkcolor=blue,citecolor=blue,filecolor=blue,urlcolor=blue,}
\usepackage{newtxtext,newtxmath}
\usepackage{graphicx,amsmath,amsfonts,amssymb}

\usepackage[T1]{fontenc}
\usepackage{ae,aecompl}

\usepackage{graphicx}	
\usepackage{amsmath}	
\usepackage{amssymb}	
\usepackage{xcolor}

\newcommand{\nhtd}{\mbox{\rm NH$_2$D(1$_{1,1}$-1$_{0,1}$)}}
\newcommand{\nthp}{\mbox{\rm N$_2$H$^+$(1-0)}}
\newcommand{\hcn}{\mbox{\rm HCN(1-0)}}
\newcommand{\hcop}{\mbox{\rm  HCO$^+$(1-0)}}

\newcommand{\htcop}{\mbox{\rm H$^{13}$CO$^{+}$(1-0)}} 
\newcommand{\methato}{\mbox{\rm CH$_3$OH(2$_{0,2}$-1$_{0,1}$)}}
\newcommand{\methatf}{\mbox{\rm CH$_3$OH(3$_{1,3}$-4$_{0,4}$)}}

\newcommand{\kms}{\,~km~s$^{-1}$}
\newcommand{\msun}{\,M$_{\odot}$}

\definecolor{cote}{HTML}{118336} 

\title[ALMA observations of FHSC candidates]{ALMA observations of envelopes
around first hydrostatic core candidates}

\author[M. Maureira et al.]{
Mar\'ia Jos\'e Maureira,$^{1,2}$\thanks{E-mail: maureira@mpe.mpg.de (MJM)} 
H\'ector G. Arce,$^{1}$
Michael M. Dunham,$^{3,4}$ 
Diego Mardones,$^{5,2}$ 
\newauthor{Andr\'es E. Guzm\'an,$^{6}$, Jaime E. Pineda,$^{2}$ and Tyler L. Bourke, $^{7,3}$}
\\
$^{1}$Astronomy Department, Yale University, New Haven, CT~06511, USA\\
$^{2}$Max-Planck-Institut für extraterrestrische Physik (MPE), Gießenbachstr. 1, D-85741 Garching, 
Germany\\
$^{3}$Center for Astrophysics | Harvard \& Smithsonian, Cambridge, MA 02138, USA\\
$^{4}$Department of Physics, State University of New York Fredonia, Fredonia, New York 14063, USA\\
$^{5}$Departamento de Astronom\'ia, Universidad de Chile, Casilla 36-D, Santiago, Chile\\
$^{6}$National Astronomical Observatory of Japan, National Institutes of Natural Sciences, 2-21-1 Osawa, Mitaka, Tokyo 181-8588, Japan\\
$^{7}$SKA Organization, Jodrell Bank, Lower Withington, Macclesfield SK11 9FT, UK\\
}

\date{Accepted XXX. Received YYY; in original form ZZZ}

\pubyear{2020}
\begin{document}
\label{firstpage}
\pagerange{\pageref{firstpage}--\pageref{lastpage}}
\maketitle

\begin{abstract}
We present  ALMA 3 mm molecular line and continuum observations with a resolution of $\sim$3.5" towards five first hydrostatic core (FHSC) candidates (L1451-mm, Per-bolo 58, Per-bolo 45, L1448-IRS2E and Cha-MMS1). Our goal is to characterize their envelopes and identify the most promising sources that could be bona fide FHSCs. We identify two candidates which are consistent with an extremely young evolutionary state (L1451-mm and Cha-MMS1), with L1451-mm being the most promising FHSC candidate. Although our envelope observations cannot rule out Cha-MMS1 as a FHSC yet, the properties of its CO outflow and SED published in recent studies are in better agreement with the predictions for a young protostar. For the remaining three sources, our observations favor a prestellar nature for Per-bolo 45 and rule out the rest as FHSC candidates. Per-bolo 58 is fully consistent with being a Class 0, while L1448 IRS2E shows no emission of high-density tracers (NH$_2$D and N$_2$H$^{+}$) at the location of the previously identified compact continuum source, which is also undetected in our observations. Thus we argue that there is no embedded source at the presumptive location of the FHSC candidate L1448 IRS2E. We propose instead, that what was thought to be emission from the presumed L1448 IRS2E outflow corresponds to outflow emission from a nearby Class 0 system, deflected by the dense ambient material. We compare the properties of the FHSC candidates studied in this work and the literature, which shows that L1451-mm appears as possibly the youngest source with a confirmed outflow.

\end{abstract}




\section{Introduction}

Low-mass stars are born out of the gravitational collapse of dense ($\sim10^{5}$ cm$^{-3}$) and cold ($\sim$10 K) prestellar cores \citep{2000AndrePrestellar}. Among the most centrally condensed prestellar cores there are some that might represent a stage shortly before the formation of a protostar, when the collapsing structure is predicted to harbor a compact central object known as a first hydrostatic core (FHSC) \citep{1969LarsonNumerical}. Although not yet observationally confirmed, the FHSC stage, lasting up to only 10$^4$ years,  is routinely seen in numerical simulations of core collapse \citep{1998Masunaga,2006SaigoEvolution,2011MatsumotoProtostellar,2012CommerconSynthetic,2013TomidaRadiation,2014BateCollapse}. The FHSC is identified as an embedded object in quasi-hydrostatic equilibrium between thermal pressure and gravity, with an initial temperature of a few 100 K and a radius of 5 to 20 au \citep{2014BateCollapse}. Due to  continuous accretion from the collapsing core, the central temperature and density of the FHSC increases during its lifetime. When the central region reaches a temperature of $\sim2000$ K, collisional dissociation of the molecular hydrogen begins, triggering a second collapse that ends with the formation of the protostar. Simulations show that the FHSC stage can play an important role on the distribution of angular momentum from the earliest stages. It is at the high-densities of the FHSC object that non-ideal MHD effects can effectively  remove angular momentum, impacting disk formation and early disk fragmentation \citep{2014MachidaConditions,2018TsukamotoMisalignment}. Additionally, the FHSC object can launch a poorly collimated and slow (few \kms) outflow before protostar formation (\citealt{2008MachidaHigh,2013TomidaRadiation,2014BateCollapse,2015TomidaRadiation}, although see \citealt{2012PriceCollimated}).\\

Observationally, a few sources have been proposed as FHSC candidates. These dense cores have extremely weak or undectected emission at wavelengths $\leq70$ $\mu$m, and thus are consistent with having an embedded FHSC or an extremely young (low-luminosity) Class 0 protostar \citep{2006BellocheEvolutionary,2010EnochCandidate,2012PezzutoHerschel,2010ChenL1448,2011PinedaEnigmatic}. Consistent with both scenarios, follow-up interferometric observations revealed compact continuum sources and small outflows ($\lesssim$ few 1,000 au lobe-size) at the center of these cores  \citep{2011DunhamDetection,2011PinedaEnigmatic,2014HiranoTwo,2015GerinNascent,2010ChenL1448,2020FujishiroLow}. A few more candidates have been been proposed based on their SED alone \citep{2018YoungWhat} and/or their compact appearance in interferometric observations of the dust emission (\citealt{2012SchneeHow,2018FriesenALMA}). For most of these new candidates the presence or absence of small outflows have yet to be determined. The most intriguing cases are recently reported in \cite{2018FriesenALMA}. They show two 1.3 mm compact sources (N6-mm and SM1N) at a resolution of $\sim$100 au, with no clear signs of an outflow. The lack of a clear outflow signature at these scales is consistent with recent theoretical predictions of the FHSC stage \citep{2019YoungSynthetic}.\\

Although both, weak or non-detected infrared emission along with compact dust and outflow detections are useful to identify candidates, they are not sufficient for identifying the youngest and most promising FHSC candidates. Molecular lines revealing the chemical and physical stage of the envelope surrounding these sources are important for identifying the candidates most likely to be bona fide FHSC. For instance, if the source is truly a young cold FHSC, or a recently-born protostar, the dense collapsing envelope (disk and pseudo-disk, if present) should show no signs of  warm ($>$20-30 K) gas/dust for which the chemistry changes due to CO desorption \citep{2004LeeEvolution}, at scales larger than about 100 au \citep{2014MachidaProtostellar,2013HincelinSurvival,2016HincelinChemical}. Similarly, the kinematics of the lines close to the center of the core should show infall and/or rotational motions in agreement with a central compact object with a very low mass ($\lesssim$ 0.1 M$_{\odot}$). 

Interferometric molecular line emission and kinematic studies have been carried out towards a few FHSC candidates. In \citealt{2017MaureiraKinematics,2017MaureiraTurbulent}, we studied the candidates L1451-mm and Per-bolo 58 at 1,000 au scales and found that they are indeed consistent with being in an extremely young evolutionary stage, either early Class 0 or FHSC. Similar results were obtained in \cite{2017FuenteChemical} for the FHSC candidates in the B1b core. More recently, \citet{2018MarcelinoALMA} found evidence of a warm inner envelope (60 K at $\sim$ 200 au) and hot corino (200 K at $\sim$100 au) towards the southern candidate in B1b (B1b-S), and thus, this source (which had been previously designated a FHSC candidate) is now confirmed as a young Class 0 protostar. On the other hand, no such hot corino signature was found towards the northern source in this system (B1b-N, also previously designated a FHSC candidate) but high outflow velocities of up to 8\kms have been revealed in new higher resolution (60 au) methanol observations \citep{2019HiranoTwo}, and is now confirmed as another very young Class 0 protostar.\\

In this work we present ALMA 3 mm molecular lines and continuum observations towards five FHSC candidates at 3.5"-4" resolution. Our goal is to further constrain the evolutionary state of these candidates by using the dust and dense gas distribution and kinematics at inner envelope scales. Based on these observations, we rule-out two candidates (L1448 IRS 2E, Per-bolo 58) and a third one shows strong evidence of being truly prestellar (Per-bolo 45). Finally, we identify one (L1451-mm) as a promising FHSC candidate for follow-up studies at high resolution.

The paper is organized as follows. In Section 2, we describe the sample and observations used in these work. In Section 3 we present our results. In Section 4 we analyze and discuss in detail the individual sources. In Section 5 we discuss the properties of the current and former FHSC candidates in this work and the literature.  We also discus future observations that can help to find new FHSC candidates as well as future high-resolution observations that can serve to confirm the true nature of the most promising targets. Section 6 corresponds to the summary and conclusions.

\section{Data}

\subsection{The sample}

\begin{table*}
\caption{Properties of the sample of FHSC candidates observed with ALMA}
\small
\label{tb:fhsc_prop}
\resizebox{\textwidth}{!}{\begin{tabular}{lccccccc}

\hline
\hline
Source &M$_{core}$(M$_{\odot}$)&L$_{int}$(L$_{\odot})$ &24$\mu$m ? $^{a}$&CO$_{out}$ lobe length$^{b}$&CO$_{out}$ velocity$^{b}$& Distance &Ref.\\
&&&  &&&&\\
\hline
\hline
L1448 IRS2E&2.3&$<0.035$&No&1114 au & 25\kms&290 pc&1\\

Per-bolo 45& 2.1& $<0.021$ & No&...&...&290 pc&2\\

Per-bolo 58&0.7&$0.020$ &Yes &7192 au & 3\kms&290 pc&3\\

L1451-mm& 0.3&$<0.021$ &No &638 au & 1\kms&290 pc&4\\

Cha-MMS1&0.6&$0.021$&Yes&2470 au & 2-12\kms&190 pc&5\\
\hline
\hline
\end{tabular}}
\vspace{1ex}
\footnotesize
\raggedright REFERENCES: 
(1) \citealt{2010ChenL1448}, (2) \citealt{2006KirkLarge,2010SadavoyMass}, (3)  \citealt{2011DunhamDetection}, (4) \citealt{2011PinedaEnigmatic} , (5) \citealt{2006BellocheEvolutionary,2019BuschDynamically}\\

\raggedright $^{a}$ Indicates whether the source has been detected at 24$\mu$m by Spitzer observations. \\
$^{b}$ Not corrected for inclination. 
\end{table*}

\begin{table*}
\caption{Maps parameters}
\label{tb:cubes_properties}
\resizebox{\textwidth}{!}
{\begin{tabular}{llcccc}
\hline
\hline
Source &Map & Rest Frequency &Synthesized Beam &PA &RMS\\
&& (GHz) & & &(mJy beam$^{-1}$)\\
\hline
\hline
\hline
L1448-IRS2E&\nhtd & 85.926& 5.4"$\times$2.8"  &-32$^{\circ}$ &24 \\
(12m + 7m)&\nthp & 93.173 &  4.2"$\times$3.0"  &17$^{\circ}$ &24 \\
&\hcn & 88.632 &  5.3"$\times$2.7"&-33$^{\circ}$ &34 \\
Phase center (J2000)&\htcop & 86.754 & 5.4"$\times$2.8" &-34$^{\circ}$ &26 \\
03h 25m 25.66s&\hcop & 89.189& 5.4"$\times$2.8" &-33$^{\circ}$  &29 \\
+30d 44m 56.70s&\methato & 96.741 & 4.1"$\times$2.9"  &19$^{\circ}$ &24 \\
&\methatf & 107.014 & 3.6"$\times$2.6"  &16$^{\circ}$ &25 \\
&3mm continuum & 103.55 & 3.5"$\times$2.5" &16$^{\circ}$  &0.15  \\
\hline
Per-bolo 45&\nhtd & 85.926 & 5.4"$\times$3.8" &-32$^{\circ}$ &34 \\ 
(12m + 7m)&\nthp & 93.173 & 4.2"$\times$3.0" &14$^{\circ}$ &21 \\ 
&\hcn & 88.632 & 5.8"$\times$2.7"&-33$^{\circ}$   &34 \\
Phase center (J2000)&\htcop & 86.754 & 5.4"$\times$2.8" &-34$^{\circ}$  &26 \\
03h 29m 07.70s &\hcop & 89.189 &5.5"$\times$2.8"&-34$^{\circ}$   &29 \\
+31d 17m 16.80s&\methato & 96.741 & 4.0"$\times$2.8"  &16$^{\circ}$ &22 \\
&\methatf & 107.014 & 3.6"$\times$2.6"  &15$^{\circ}$ &24 \\
&3mm continuum & 103.55 & 3.5"$\times$2.5"  &13$^{\circ}$ &0.15 \\

\hline
Per-bolo 58&\nhtd & 85.926 & 5.5"$\times$2.8" &-32$^{\circ}$ &38 \\ 
(12m + 7m)&\nthp & 93.173 & 4.2"$\times$3.0" &17$^{\circ}$ &21 \\ 
&\hcn & 88.632 & 5.4"$\times$2.7" &-34$^{\circ}$ &34 \\
Phase center (J2000)&\htcop & 86.754 & 5.5"$\times$2.8"  &-34$^{\circ}$  &26 \\
03h 29m 25.46s&\hcop & 89.189 & 5.5"$\times$2.8" &-33$^{\circ}$&29 \\
+31d 28m 15.00s&\methato & 96.741 & 4.1"$\times$2.8" &17$^{\circ}$ &22 \\
&\methatf & 107.014 &  3.6"$\times$2.6" &15$^{\circ}$ &25 \\
&3mm continuum & 103.55 & 3.5"$\times$2.4" &14$^{\circ}$ &0.15 \\
\hline

L1451-mm&\nhtd & 85.926 & 5.4"$\times$2.8" &-33$^{\circ}$ &34 \\ 
(12m + 7m)&\nthp & 93.173 & 4.1"$\times$3.0" &16$^{\circ}$ &20 \\ 
&\hcn & 88.632  & 5.3"$\times$2.7"  &-34$^{\circ}$  &34 \\
Phase center (J2000)&\htcop & 86.754 & 5.4"$\times$2.8" &-34$^{\circ}$ &26 \\
03h 25m 10.21s&\hcop & 89.189&  5.4"$\times$2.8" &-34$^{\circ}$  &29 \\
30d 23m 55.30s&\methato & 96.741 & 4.0"$\times$2.8" &17$^{\circ}$ &21 \\ 
&\methatf & 107.014 & 3.6"$\times$2.6" &14$^{\circ}$ &25 \\ 
&3mm continuum & 103.55 & 3.5"$\times$2.5" &14$^{\circ}$ &0.14 \\ 
\hline
Cha-MMS1&\nhtd & 85.926 & 3.1"$\times$1.8" &11$^{\circ}$ &14 \\
(12m + 7m + TP)&\nthp & 93.173 & 2.8"$\times$1.8" &-11$^{\circ}$  &20 \\
&\hcn & 88.632 & 3.0"$\times$1.8" &11$^{\circ}$ &13 \\
Phase center (J2000)&\htcop & 86.754 & 3.1"$\times$1.8"&12$^{\circ}$&15 \\
11h 06m 31.23s&\hcop & 89.189& 3.0"$\times$1.8" &12$^{\circ}$&14 \\
-77d 23m 33.57s&\methato & 96.741 & 2.7"$\times$1.8" &-11$^{\circ}$ &22 \\
&\methatf & 107.014 & 2.5"$\times$1.6" &-10$^{\circ}$ &27 \\
&3mm cont (12m+7m only) & 103.55 & 2.5"$\times$1.6" &-7$^{\circ}$  &0.13  \\

\hline
\hline
\end{tabular}}
\vspace{1ex}
\raggedright Notes: the RMS for the molecular lines is measured using channels that are 0.1 \kms\ wide.
\end{table*}

The sample of sources targeted by our ALMA observations included all FHSC candidates proposed in the literature up to the year 2012 (when ALMA Cycle 1 proposals were due) that are observable by this southern facility. This corresponded to five sources: L1448 IRS2E, Per-bolo 45, Per-bolo 58, L1451-mm and Cha-MMS1. 

The observed sources in this study are very low luminosity objects ($<0.01-0.1$ L$_{\odot}$).  Only two have been detected at 24 $\mu$m and 70 $\mu$m (Per-bolo 58 and Cha-MMS1), with fluxes comparable to the upper limits of the remaining undetected candidates. All five sources have been associated with a far-IR/sub-mm core in low-resolution (single-dish) observation as well as with a compact sub-mm/mm continuum source in interferometric observations.  Interferometric CO line observations revealed an associated molecular outflow in L1448 IRS2E \citep{2010ChenL1448}, Per-bolo 58 \citep{2011DunhamDetection}, L1451-mm
\citep{2011PinedaEnigmatic}, and Cha-MMS1 (\citealt{2019BuschDynamically}), while in Per-bolo 45 there is no clear outflow detection (\citealt{2012SchneeHow,2019StephensMass}). \\

\noindent Table~\ref{tb:fhsc_prop} lists general properties of the sample of FHSC candidates observed with our ALMA cycle 1 observation. We have updated the values of the properties that depend on the distance to the source (core mass, internal luminosity, outflow lobe length) to be consistent with the latest distance estimates to the Perseus and Chamaeleon molecular clouds.  The adopted distances are 290 pc (\citealt{2018ZuckerMapping,2018OrtizLeonGould}) for the sources in Perseus (L1448 IRS2E, Per-bolo 45, Per-bolo 58 and L1451-mm) and 190 pc for Cha-MMS1 (\citealt{2018A&ARoccatagliataDouble,2018DzibDistances}).

\subsection{Observations}

Observations for all sources in our sample were conducted with the ALMA 12m array and the Atacama Compact Array (ACA). For Cha-MMS1, observations with the Total Power (TP) array were also obtained. The ALMA cycle 1 Observations with the 12m and 7m arrays were taken between December 2013 and August 2014. TP array observations were taken on April 2015. Each source was observed using a 12m array 7-pointing mosaic and a 3-point mosaic with the 7m array. The projected baselines ranged between 7 m and 1290 m, resulting in a largest recoverable scale of about 60" and a resolution of 3.5" to 4".

\noindent The observations (all taken under project code 2012.1.00394.S, PI D.~Mardones) included two different frequency setups in Band 3. One setup consisted of four spectral windows which included the lines  \nhtd, \hcn, \htcop\ and \hcop. The other setup consisted of three  windows that included the lines \nthp, \methato, \methatf, and a window for continuum observations. The bandwidths for the line and continuum windows were 58.6 MHz and 2000 MHz, respectively. The corresponding channel spacings for these windows were  30.52 KHz ($\sim$0.1\kms) and 31250.00 kHz ($\sim$88\kms).\\  

\noindent The Common Astronomy Software Applications (CASA) package, was used for both calibration and imaging. Calibration of the raw visibility data was done using the standard reduction script provided separately for the 12m, 7m and TP array observations. Emission-free channels in all the spectral windows were combined with the continuum band to create the continuum image, which resulted in a continuum bandwidth of 2438 MHz. Imaging of the combined 12m and 7m data was done using the task {\it tclean}  with a Briggs robust parameter of $0.5$. We used the multiscale option with a threshold for cleaning of $2-3\sigma$. 

For Cha-MMS1, the combined 12m and 7m CLEANed image was combined with the TP image using the task {\it feather}. The size of the final synthesized beams for all maps are about 4", which at the distance of the Perseus molecular cloud (290 pc) and  the Chamaeleon I molecular cloud (190 pc) corresponds to about 1,200 au and 800 au, respectively. Table~\ref{tb:cubes_properties} lists the line frequencies, beam sizes and RMS of the final data cubes.

\subsection{ALMA archive data}

We calibrated and imaged  ALMA Cycle 2 (band 6) archival single pointing observations towards Cha-MMS1 (2013.1.01113.S, PI: Cordiner). We used CASA for both calibration (using the standard provided pipeline) and imaging. We used the continuum window and all available line-free channels for producing the final continuum image. We further improved the calibration by performing self-calibration on the continuum. The final image was generated with a robust parameter of 0.5, using multiscale cleaning.
The final beam size, beam P.A and noise of the continuum image are 0.73"$\times$0.41", -5.6$^{\circ}$ and 0.025 mJy/beam, respectively. We applied the final self-calibration solution to the CO(2-1) and CS(5-4) line windows, and produced cubes for both lines using a robust parameter of 2, with the multiscale deconvolver option. The final beam sizes and beam P.A. are 0.77"$\times$0.51", 0.73"$\times$0.49" and -33$^{\circ}$,-32$^{\circ}$ for CO and CS, respectively. The noise, per channel, of the cubes are 5.7 mJy/beam and 5.5 mJy/beam for CO and CS, respectively.

\section{Results}

\subsection{3mm Continuum}

\begin{table*}

\caption{Properties of Continuum Sources from Gaussian Fit}
\label{tb:cont_fit}
\begin{tabular}{lcccccc}
\hline
\hline
Source &R.A. & Decl. &Deconvolved Size &PA &Peak Flux Density&Integrated Flux Density\\
&(J2000)& (J2000) &(arcsec) & $\deg$ &(mJy beam$^{-1}$)&(mJy)\\
\hline
\hline
L1448-IRS2E&... &...& ... & ...
&<0.15 &<0.15\\

Per-bolo 45&03:29:06.97 &+31.17.22.43& (13.49 $\pm$ 3.01 , 3.40 $\pm$ 0.83) & 3.2$^{\circ}$($\pm$4.4)
&0.63 $\pm$ 0.11&4.31 $\pm$ 0.86 \\

Per-bolo 58&03:29:25.49 &+31:28:15.33& (10.05 $\pm$ 1.61 , 4.42 $\pm$ 0.73) & 28.6$^{\circ}$($\pm$7.1)
&0.95 $\pm$ 0.12&6.02 $\pm$ 0.87 \\

L1451-mm&03:25:10.24 & +30:23:55.02& point source &...
&4.00 $\pm$ 0.14&4.17 $\pm$ 0.25 \\

Cha-MMS1&11:06:33.46 & -77:23:34.52& (3.15 $\pm$ 0.49 , 1.35 $\pm$ 0.42) & 142$^{\circ}$($\pm$11)
&1.74 $\pm$ 0.13&3.94 $\pm$ 0.42 \\
\hline
\hline
\end{tabular}
\end{table*}

\begin{figure*}
   \includegraphics[width=1\textwidth]{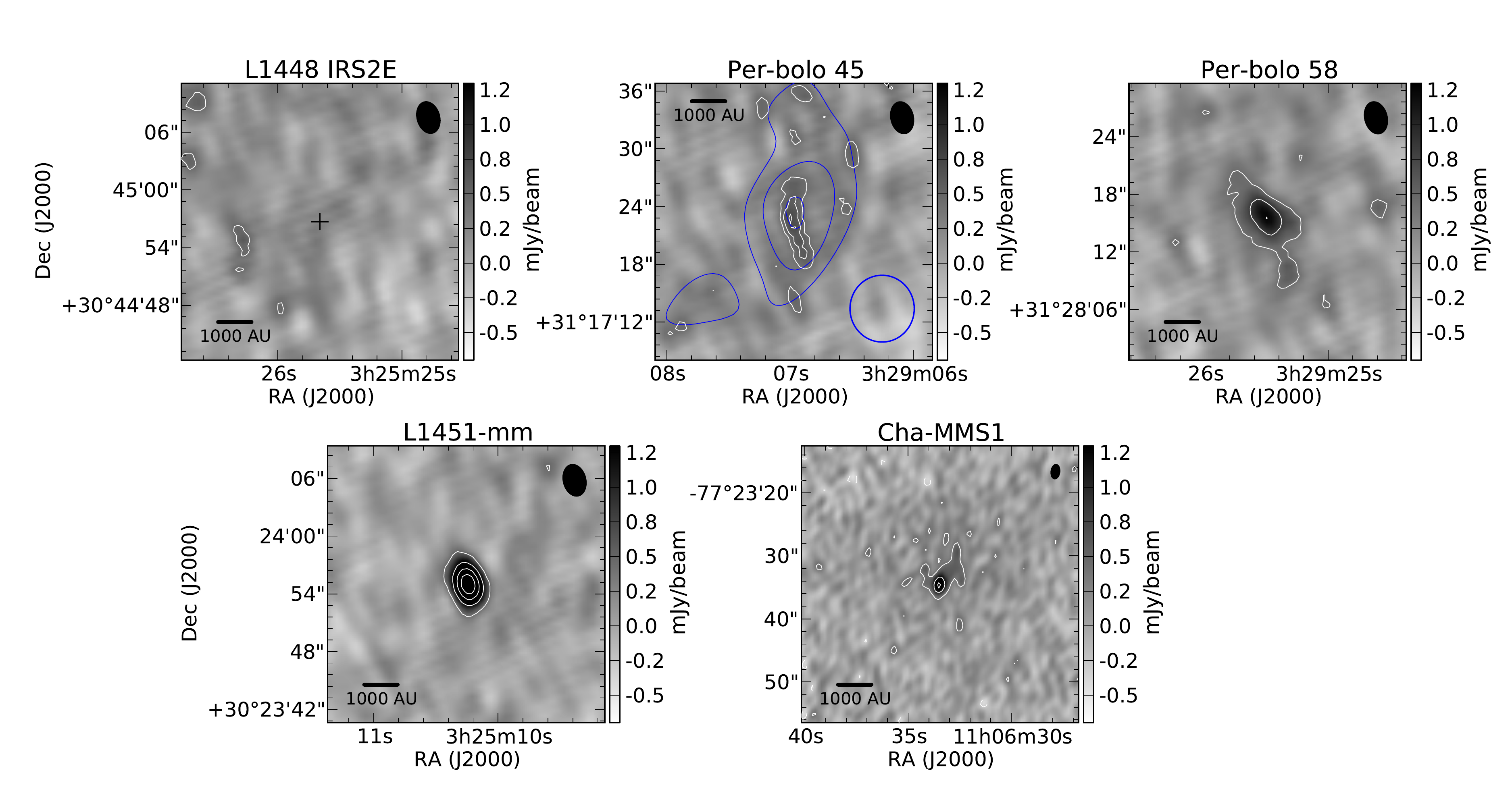}
    \caption{3 mm continuum for all the observed sources (color and white contours). White contours start at 3$\sigma$ for all sources, and increase in steps of 1$\sigma$, 3$\sigma$, 6$\sigma$ and 6 $\sigma$ for Per-bolo 45, Per-bolo 58, L1451-mm and Cha-MMS1, respectively. The location of the 1.3 mm SMA continuum peak for L1448 IRS2E \citep{2010ChenL1448} is marked with a cross. The synthesized beam for each map is shown in black. For Per-bolo 45, we also show the 3 mm continuum emission imaged with a beam of  $\sim$7" (blue contours), in order to compare with previous CARMA observations of this source (\citealt{2012SchneeHow}). 
    \label{fig:cont_all}}
\end{figure*}

Figure~\ref{fig:cont_all} shows the 3 mm continuum of all the observed sources. We detected all but the FHSC candidate L1448 IRS2E. For the later, we expected a detection of at least $2\sigma$, calculated by extrapolating the $\sim$6 mJy point source detection at 1.3 mm with the SMA \citep{2010ChenL1448}, and assuming a dust opacity spectral index $\beta\approx1.7$. 

For the detected sources, Per-bolo 45 has the weakest peak flux density and also shows the less concentrated emission, which appears to be almost resolved out. The rest are point-like sources, some of them also showing extended emission.

\begin{table}

\caption{Physical Properties of the Dust Continuum Emission}
\label{tb:physical_fit}
\begin{tabular}{lccc}
\hline
\hline
Source &Effective Radius&Mass &n\\
&au& (M$_{\odot}$) &(cm$^{-3}$)\\
\hline
\hline
L1448-IRS2E& ... &$<$0.01 &...\\

Per-bolo 45& 982(162)&0.29(0.06) &$1.1(0.6)\times10^{7}$\\

Per-bolo 58&966(111)&0.41(0.06) &$1.6(0.6)\times10^{7}$ \\

L1451-mm&$<429$ & 0.28(0.02) &$>1.3\times10^{8}$ \\

Cha-MMS1&196(34)  &0.12(0.01)&$5.4(3.5)\times10^{8}$ \\

\hline
\hline
\end{tabular}
\vspace{1ex}
\footnotesize
\raggedright Values in parenthesis correspond to the uncertainties. \\

\end{table}

Using CASA task {\it imfit} we fit a single elliptical Gaussian component to the 3mm continuum emission for the detected sources. This provides  the position, deconvolved size, peak flux density and total flux for each source, which are listed in Table~\ref{tb:cont_fit}. Table~\ref{tb:physical_fit} uses these values to estimate sizes in au, masses and total gas number density. Masses from the 3 mm continuum are calculated using:
\begin{align}
M=\frac{d^2S_{\nu}}{B_{\nu}(T_{d})\kappa_{\nu}}
\end{align}
\noindent where $S_{\nu}$ is the integrated flux density, $B_{\nu}$ is the Planck function,   and $\kappa_{\nu}$ is the dust mass opacity. We assume that  the dust temperature, $T_{D}$, has a value of 10 K, suitable for a prestellar core, or a core with a low luminosity object. We adopt a dust opacity of 0.23 cm$^2$g$^{-1}$ at 104 GHz, corresponding to an extrapolation, using $\beta=1.7$ \citep{2016ChenJCMT}, of the 1.3 mm value in \cite{1994OssenkopfDust} for thin ice mantles after $10^5$ years of coagulation at a gas density of 10$^6$ cm$^{-3}$ (so-called OH5 dust). The mass values in Table~\ref{tb:physical_fit} are calculated assuming a gas-to-dust ratio of 100. The effective radius $r_{eff}$ and total gas number density $n$ were calculated following the convention in \cite{2016DunhamALMA}. The errors in Table~\ref{tb:physical_fit} only consider the errors from the elliptical Gaussian fit (Table~\ref{tb:cont_fit}). 
\noindent Masses between 0.12 M$_{\odot}$ and 0.41 M$_{\odot}$ and densities between $1.1\times10^7$ cm$^{-3}$ and $5.4\times10^8$ cm$^{-3}$ were estimated for the detected candidates. The uncertainties in the masses and thus also gas number density are larger than those reported in Table~\ref{tb:physical_fit} (up to a factor of a few) due to the unaccounted uncertainties in the dust temperature, dust opacity, and distance. For instance, assuming a different type of dust with a higher opacity (e.g., \citealt{2015YenObservations}) would decrease the estimated masses by a factor of two to three. Further, if we were to assume a value of $\beta$ that is closer to 1 \citep{2017BraccoProbing}, it would also increase the opacity value at 3 mm and thus would result in mass estimates that are about twice as small as our original estimate. Finally, slightly higher temperatures due to internal heating would also decrease our estimated masses. For example, a temperature of 12 K would result in mass estimates that are approximately 20\% lower. All these factors would result in  reduction of our estimated total mass using the 3 mm dust emission up to a factor of 6, thus in agreement with the estimates of the central object mass obtained through the gas kinematics in Section~\ref{sec:analysis_per_source}.

\subsection{Molecular lines}

\begin{figure*}
\includegraphics[width=0.85\textwidth]{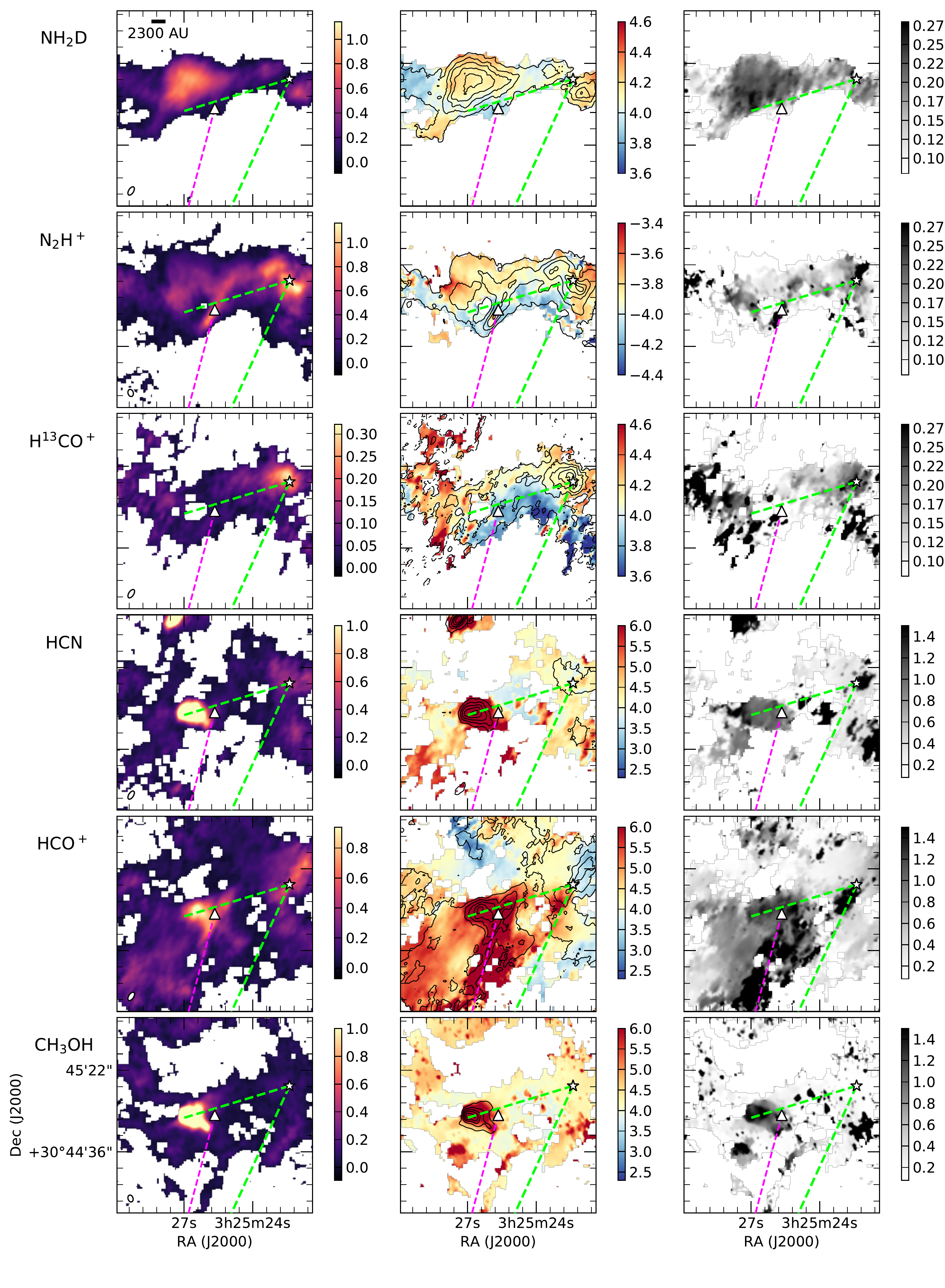}
    \caption{ {\bf L1448 IRS2E}. Moment 0 (left), moment 1 (middle) and moment 2 (right) maps for all the molecules with detected emission. The dashed green lines mark the positions of the cavity walls of the outflow driven by the nearby L1448 IRS2 (Class 0) protostar. The magenta dashed  line marks the direction of the red-shifted outflow lobe from L144 IRS2E proposed by \protect\cite{2010ChenL1448}. The positions of L1448 IRS2E and L1448 IRS2 are marked with a white triangle and a white star symbols, respectively.  Black contours following the moment 0 are overlaid on the moment 1 map (middle). Contours start at 10\% of the maximum and increase in steps of 20\%. Color bars are in units of Jy beam$^{-1}$\kms, \kms\ and \kms\ for moment 0, 1 and 2, respectively. For each molecule, the synthesized beam is shown as a black ellipse (filled-in white) at the lower left corner of the moment 0 panel. 
        \label{fig:irs2e_moments}}
\end{figure*}

\begin{figure*}
\includegraphics[width=0.9\textwidth]{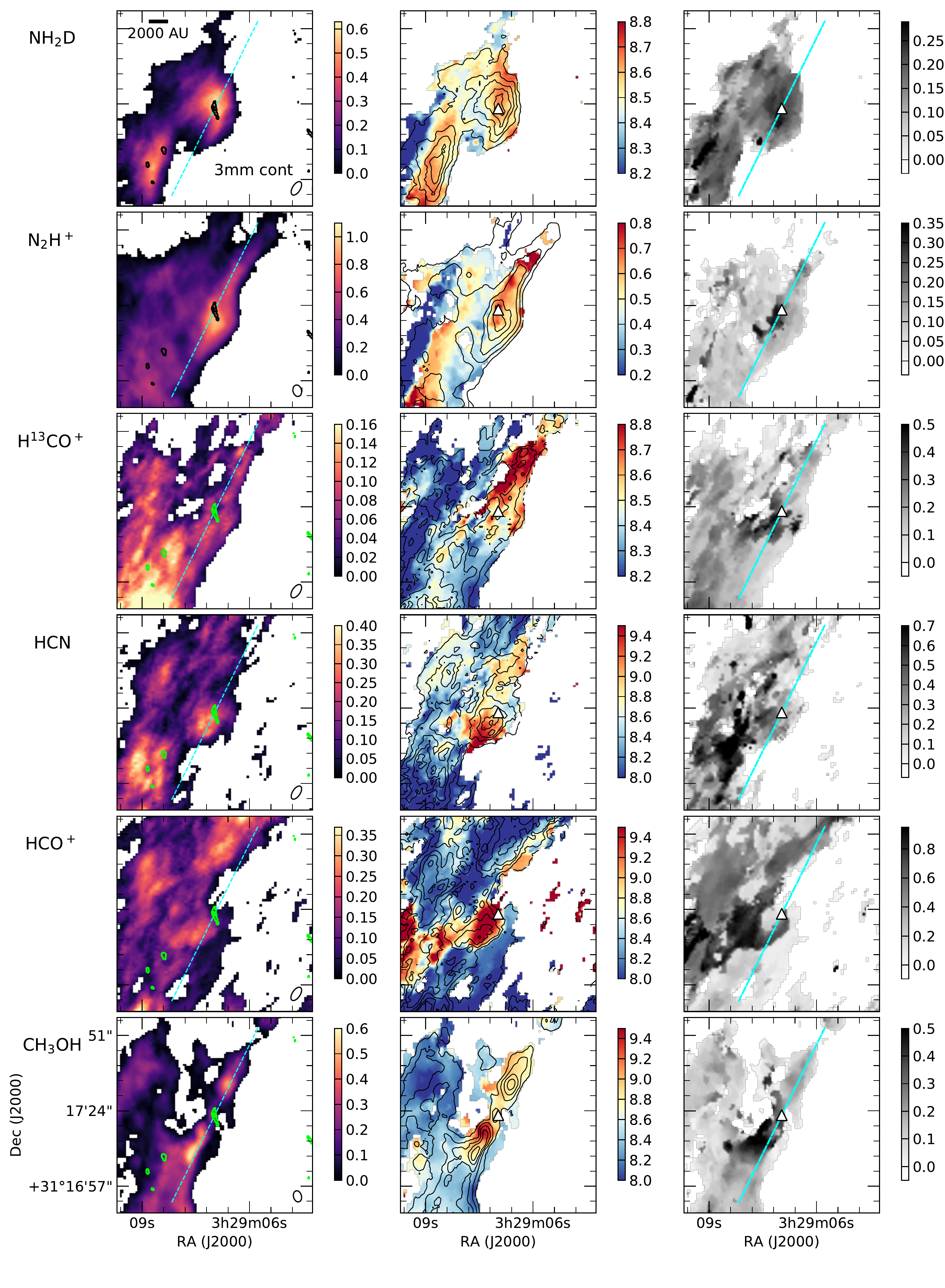}
    \caption{ {\bf Per-bolo 45}. Similar to  Figure~\ref{fig:irs2e_moments}, but for Per-bolo 45. The 3 mm continuum is shown in black/green contours on the moment 0, while the peak continuum emission is marked with a white triangle in the moment 1 and 2 columns. The cyan dashed-line shows the direction of the position-velocity cut in Figure~\ref{fig:pvmaps_per45obs_out}.
    \label{fig:perbol45_moments}}
\end{figure*}

\begin{figure*}

	\includegraphics[width=0.9\textwidth]{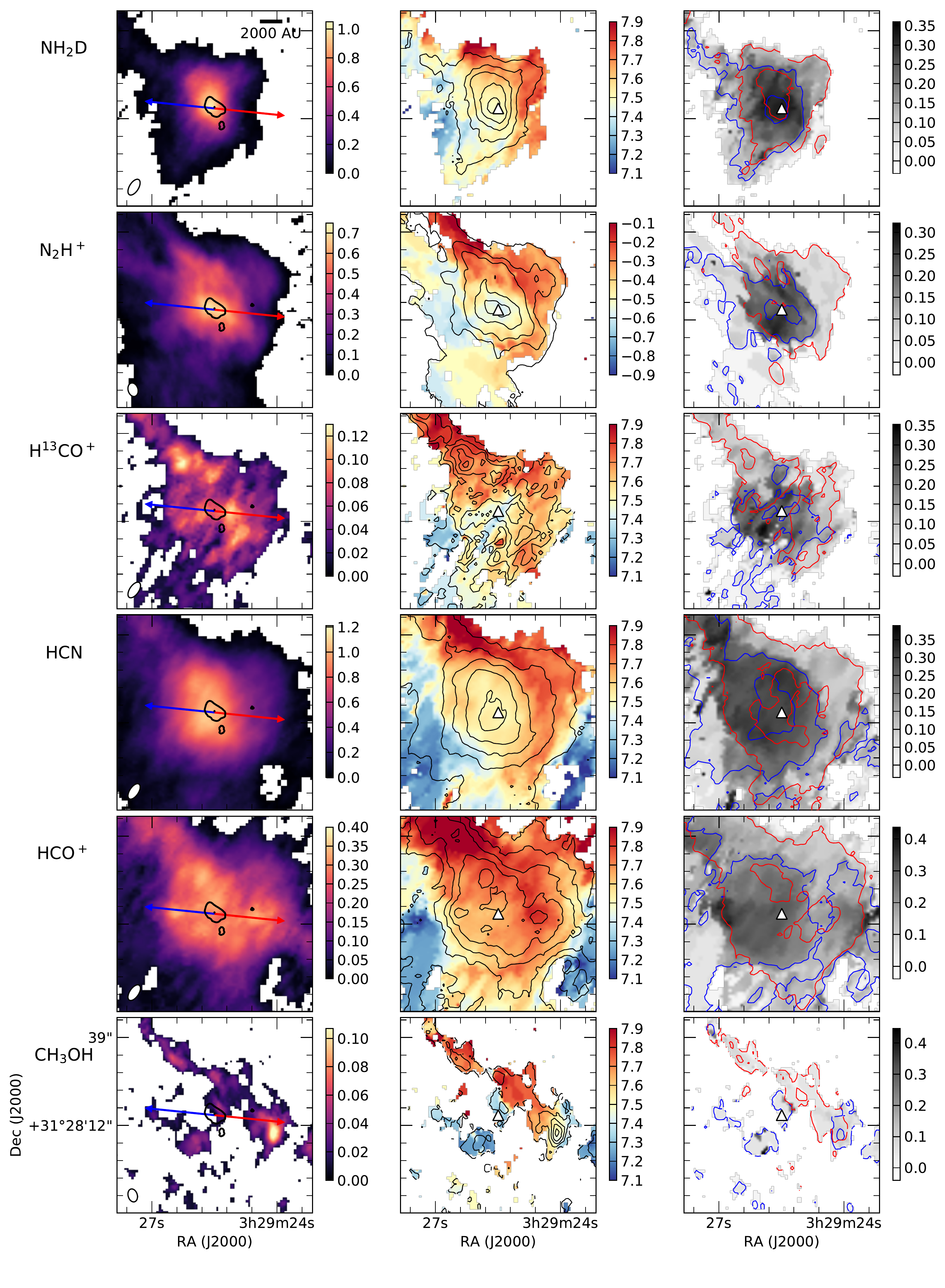}
    \caption{ {\bf Per-bolo 58}. Similar to  Figure~\ref{fig:irs2e_moments}, but for Per-bolo 58. The 3 mm continuum is shown in black contours on the moment 0, while the peak is marked with a white triangle in the moment 1 and 2 maps (middle and right columns, respectively). Red and blue contours on the moment 2 (right) show the integrated intensity of the blue velocity component (7.1-7.5\kms) and the red velocity component (7.6-8.1\kms), corresponding to the two distinct velocity components seen in the spectra (Figure~\ref{fig:per58_spec}). Blue and red contours are plot at the 20\% and 80\% of the maximum. The blue and red arrow show the direction and extent of the blue and red lobe of this source's outflow. 
    \label{fig:perbol58_moments}}
\end{figure*}

\begin{figure*}
	\includegraphics[width=0.9\textwidth]{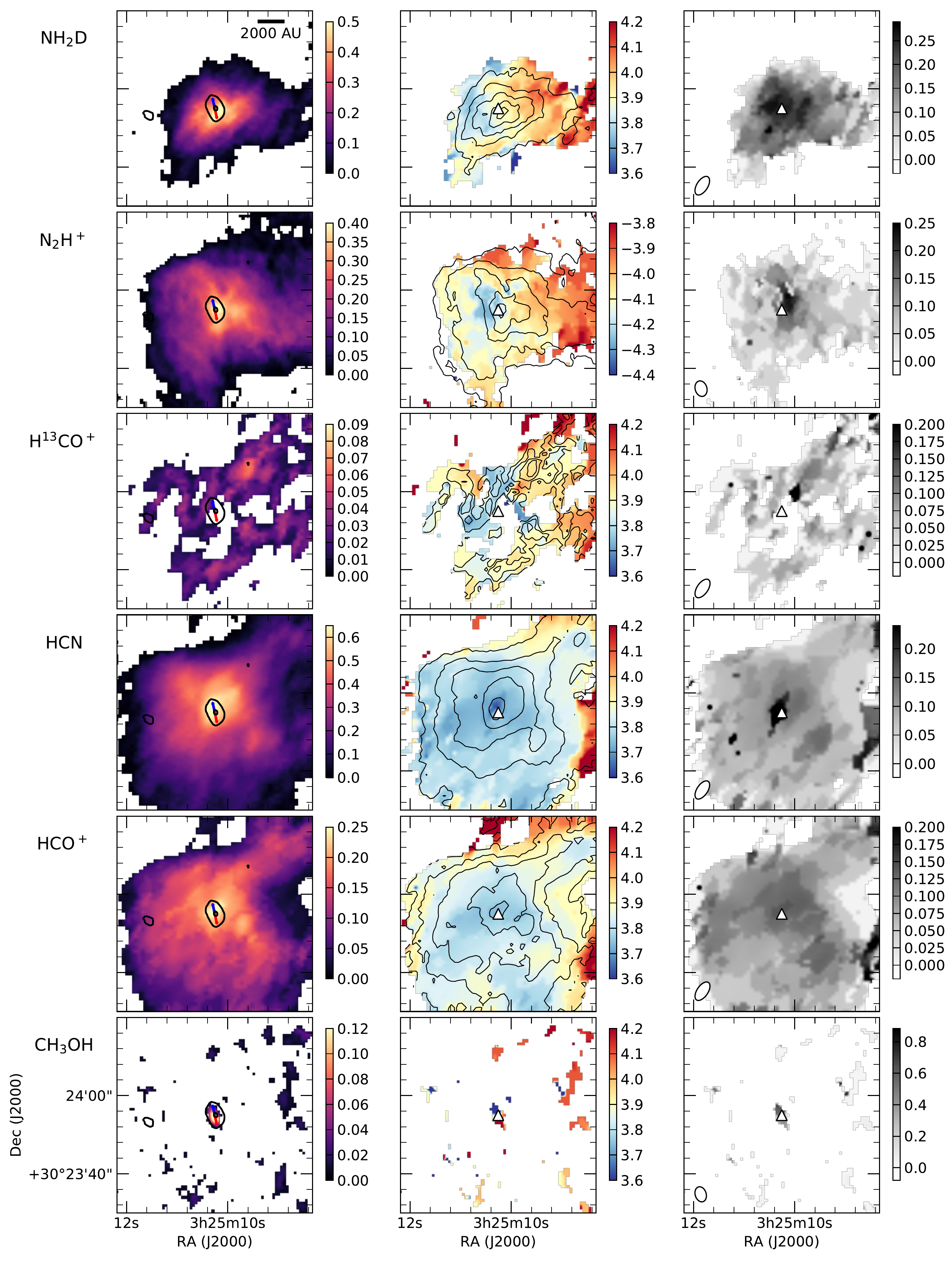}
    \caption{{\bf L1451-mm}. Similar to Figure~\ref{fig:irs2e_moments}, but for L1451-mm. The 3 mm continuum is shown in black contours on the moment 0, while the peak is marked with a white triangle in the moment 1 and 2 maps, in the middle and right columns, respectively. The blue and red arrow show the direction and extent of the blue and red lobe of this source's outflow. \label{fig:l1451mm_moments}}
\end{figure*}

\begin{figure*}
\includegraphics[width=0.9\textwidth]{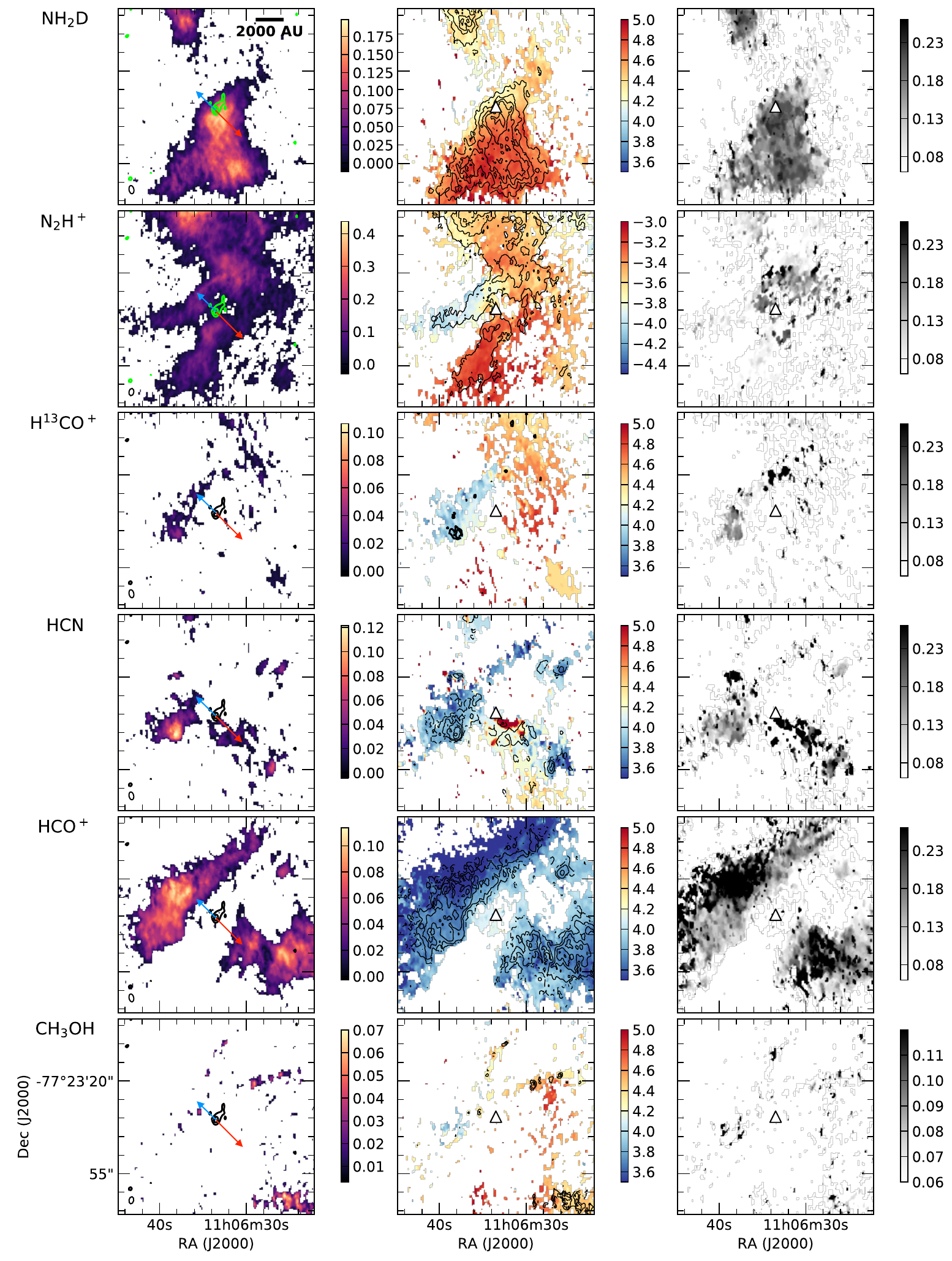} 
    \caption{ {\bf Cha-MMS1}. Similar to Figure~\ref{fig:irs2e_moments}, but for Cha-MMS1. The 3 mm continuum is shown in black/green contours on the moment 0 maps (starting at 4$\sigma$ and increasing in steps of 6$\sigma$), while the peak is marked with a white triangle in the moment 1 and 2 maps (middle and right columns, respectively). The blue and red arrow show the direction and approximate extent of the blue and red lobes of this source's outflow. These maps have not been combined with the Total Power data (see Figure~\ref{fig:cham_co32_ladd} for the final combined 12m+7m+TP maps). 
    \label{fig:cham_moments}}
\end{figure*}

\noindent Figures \ref{fig:irs2e_moments} to~\ref{fig:cham_moments} show the moment 0, 1 and 2 maps for all the observed sources and molecular lines. Moment 0 maps (left column) show integration over all velocities, but only pixels with emission greater than 3$\sigma$ per channel were considered. Moment 1 and 2 (middle and right column, respectively) were calculated using the main hyperfine for \nhtd\ and \hcn, while the isolated hyperfine component was used for \nthp\ (hence the difference in the absolute velocity shown for this line). The kinematics and emission properties of each individual source are discussed in detail in Section~\ref{sec:analysis_per_source}.

\subsubsection{General trends in the molecular line emission}
\label{sec:line_trends}

In this section we describe general trends seen in our observed molecular lines maps, to be used later in the analysis and discussion of individual sources (Section~\ref{sec:analysis_per_source}). \\

\noindent \emph{NH$_2$D (1$_{1,1}$-1$_{0,1}$):} this line is the highest density tracer in our sample. The integrated intensity of this line shows the most compact morphology for all the sources. The contour at 90\% of the maximum encloses the location of the continuum peak in all sources detected in the continuum, except for Cha-MMS1. In general, the emission shows a flattened morphology, with the major axis oriented perpendicular to the outflow (when present). \\

\noindent \emph{N$_2$H$^+$(1-0):} this line has an effective excitation density of $1.5\times10^{4}$ cm$^{-3}$ at T$=10$ K, thus it is also a high-density tracer. Similar to \nhtd, the contour at 90\% of the maximum of the integrated intensity encloses the location of the continuum peak
in all compact sources detected in the continuum, except for Cha-MMS1. However, for these sources the morphology of the emission does not trace the same flattened morphology as the \nhtd. This seems to indicate that the \nhtd\ is a better tracer of the flattened inner envelope region
compared to \nthp. \\

\noindent \emph{H$^{13}$CO$^+$(1-0):} This line is the most optically thin among the intermediate to low gas density tracers of the envelope in our sample (H$^{13}$CO$^+$, HCN, and HCO$^+$). The emission is generally weak and does not show a prominent peak near or around the continuum location in all cases, except for Per-bolo 58. \\

\noindent\emph{HCN(1-0), HCO$^+$(1-0), CH$_3$OH (2$_{0,2}$-1$_{0,1}$):} These lines trace envelope, outflow material and/or envelope outflow interactions with no clear trends in their emission distribution.\\

\section{Analysis and discussion of individual sources}
\label{sec:analysis_per_source}

In the following we discuss previous observations as well as our results and interpretation of the molecular emission towards each of the candidates.

\subsection{L1448 IRS2E: star-forming core or dense gas interacting with neighboring outflow?}

L1448 IRS2E, located in the L1448 region in Perseus, was proposed as a FHSC candidate by \cite{2010ChenL1448}. This was based on the detection of an unresolved 1.3 mm continuum source with SMA observations at a resolution of $\sim$1,000 au, located at the northern end of a high-velocity CO(2-1) jet-like structure. The presumed outflow emission 
extends $\sim$9,000 au to the south, and it is only seen at red-shifted velocities (dashed magenta line in Figure~\ref{fig:irs2e_moments}). 

Our observations in Figure~\ref{fig:irs2e_moments} show that the 1.3mm SMA point source lies near the northern cavity wall of the outflow driven by the nearby Class 0 system L1448 IRS2. Additionally, the redshifted lobe of the FHSC candidate, as proposed in \cite{2010ChenL1448}, is inside the cavity of the redshifted outflow from L1448 IRS2. \\

Many of the trends seen in the line emission maps in the other sources
discussed in section~\ref{sec:line_trends} are  not seen towards this FHSC candidate. For instance, there is no \nhtd\ emission surrounding the location of the 1.3mm SMA point source; instead, this line traces a clump located 6,000 au North-East of the proposed candidate location. Close to the proposed location of this candidate, the \nthp\ traces a narrow and elongated structure which frames the eastern side of the CO red-shifted emission observed by \cite{2010ChenL1448} that was proposed as the first core candidate's outflow.

\begin{figure}
\centering
\includegraphics[width=0.5\textwidth]{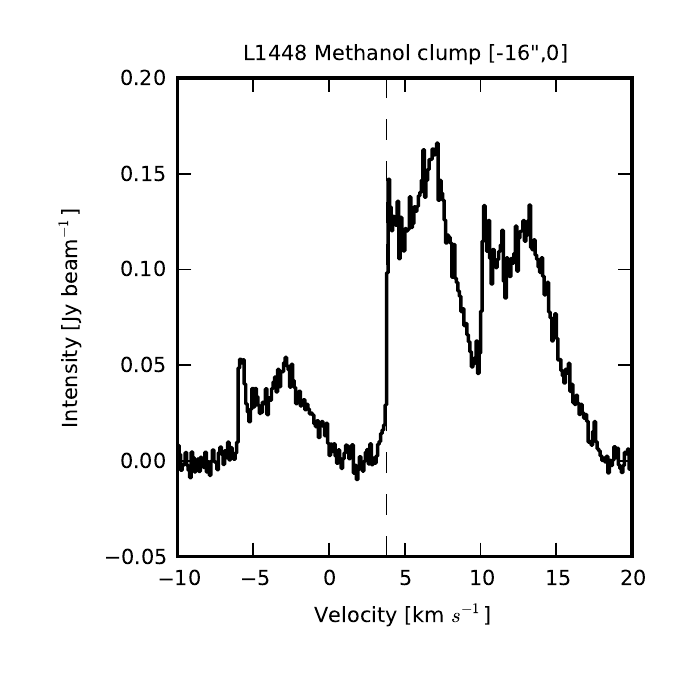}
    \caption{ \methato\ spectrum towards the bright clump to the East of the L1448 IRS2E candidate location. The spectrum is averaged over a circle with radius 10", enclosing most of the clump. This type of broad line showing a sharp edge and an extended tail of high velocities is typical of shocked regions. The vertical line corresponds to $v=3.8$ \kms, the systemic velocity at the location of the candidate (see spectra  Figure~\ref{fig:irs2e_spec}).
    \label{fig:irs2e_spec_methanol}}
\end{figure}

The \hcn, \hcop\ and \methato\ integrated intensity maps peak in a compact clump, located $\sim$5,000 au East of the 1.3 mm SMA compact source. The location of this clump is coincident with the region where the northern edge (cavity wall) of the neighboring L1448 IRS2 outflow appears to be interacting with the surrounding dense gas. The \methato\ spectrum averaged over this region shows a line with a sharp edge at the systemic velocity measured at the presumed location of the candidate ($v = 3.8$~km~s$^{-1}$) followed by relatively bright redshifted emission extending more than 10~\kms\ from the systemic velocity (see Figure~\ref{fig:irs2e_spec_methanol}). This type of line is typically seen towards shocked regions impacted by protostellar outflows \citep{2014SuutarinenWater}. Similar wide velocity profiles are seen towards this clump in \hcn\ and \hcop. In addition, the velocity range of this gas is consistent with the velocity range of the red-shifted outflow lobe from L1448 IRS2 \citep{2010ChenL1448}.\\

\noindent Our observations thus indicate a lack of concentrated emission from high-density tracers at the location of the proposed candidate, along with evidence of the impact of the neighboring L1448 IRS2 outflow near the presumed  first core candidate location, traced by low-density and shock gas tracers. Based on these, and the lack of detection of a 3mm continuum point source in our ALMA observations, we propose that the red-shifted emission interpreted as outflow from a FHSC candidate by \cite{2010ChenL1448} is emission from the outflow from the nearby Class 0 system L1448 IRS2 that has been deflected by the dense gas present at the location where L1448 IRS2E was thought to be. Thus, we believe that there is no star-forming core at this location and rule-out L1448 IRS2E as a FHSC. This interpretation is also consistent with the recently reported non-detection of 1.3 mm continuum emission in the SMA observations with a resolution of 300 au, as part of the MASSES survey \citep{2019StephensMass}. Those observations had enough sensitivity to detect the $\sim$6 mJy point source from \cite{2010ChenL1448} with a significance of 8$\sigma$.

\subsection{Per-bolo 45: a prestellar core?}
\label{sec:per_45}
\begin{figure*}
\includegraphics[width=1\textwidth]{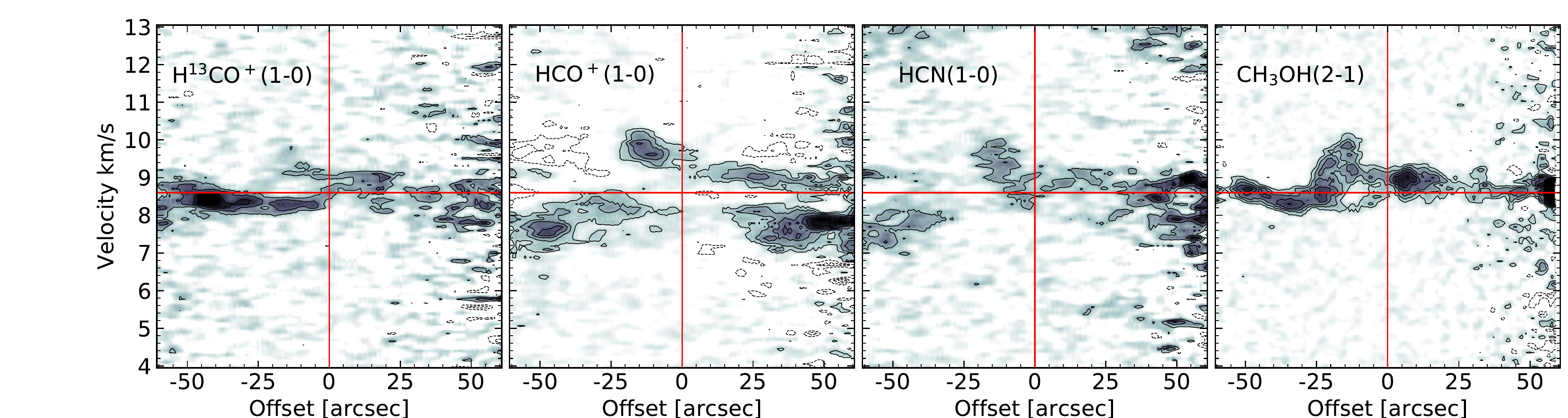}
    \caption[Per-bolo 45 position-velocity diagrams for several molecules]{ Per-bolo 45 position-velocity diagrams along the \methato\ elongated structure (PA$_{out}=-27^{\circ}$), shown in Figure~\ref{fig:perbol45_moments}. The \htcop\ is an envelope tracer while the rest are used to probe outflow motions. We do not find clear evidence for an outflow driven by this candidate. Contours start at 3$\sigma$ and increase in steps of 3.5$\sigma$ ($\sigma$ values listed in  Table~\ref{tb:cubes_properties}). Similar dashed contours are drawn for negative emission. Red lines mark the position of the continuum source and the source velocity $v=8.6$\kms.
    \label{fig:pvmaps_per45obs_out}}
\end{figure*}

Per-bolo 45 is a bound  dense Far-IR/sub-mm core \citep{2008EnochMass} located in the Perseus NGC 1333 cluster region, in between IRAS 7 to the North-East and SVS 13 to the South-West. Single-dish observations (with a FWHM resolution of 30"), along with Spitzer observations were used to classify Per-bolo 45 as a starless core \citep{2008EnochMass}. Subsequent  
interferometer observations with a resolution of $\sim$7" lead \citet{2012SchneeHow}
to propose this source as a FHSC candidate. Their observations detected a compact ($\sim$3,300 au) continuum source with molecular lines showing broader linewidths towards the center. In addition, they detected  extended SiO(2-1) emission, 20" South from the continuum source, with an elongated morphology somewhat reminiscent of an outflow (yet with velocities within 1\kms\ of the ambient velocity). They interpreted these results as tentative evidence for an outflow and concluded that Per-Bolo 45 might already had formed a compact central object, such as FHSC or a young protostar. \\

In our observations, the \nhtd\ and \nthp\ emission concentrate around the weak continuum emission. On the other hand, the \htcop\ shows lower emission at the position of the continuum source, within a region with a size of $\sim$1,000 au, which could be explained by freeze-out of carbon-bearing molecules \citep{2004LeeEvolution,2007berginCold}. These features are consistent with a dense core in a young evolutionary stage (prestellar to young Class 0 protostar). Yet, the weak elongated dust structure at 1,000 au scales (Figure~\ref{fig:cont_all}) is more consistent with what one would expect for a prestellar (or starless) core \citep{2016DunhamALMA}. To further favor or reject the prestellar nature of this core, we investigate whether our observations support the presence of an outflow.

\subsubsection{Searching for an outflow powered by Per-bolo 45}

Consistent with the observations in \cite{2012SchneeHow}, Figure~\ref{fig:irs2e_moments} shows that the emission and linewidths of the \methato, \hcop, and \hcn\ lines indeed suggest that this source might have an outflow. They show an elongated structure at red-shifted velocities to the south of the FHSC candidate with a corresponding broadening of the lines. The SiO(2-1) detected in \cite{2012SchneeHow} is about 10$\arcsec$ south from Per-bolo 45 and is oriented along the same direction as our observed  \methato\ emission, at similar velocities. Yet, the velocity of these features remains within $\sim1.5$ \kms\ from the source systemic velocity, i.e., they do not trace truly high-velocity gas.

We made position-velocity diagrams of these lines to investigate if these features could correspond to a low-velocity outflow.  Figure~\ref{fig:pvmaps_per45obs_out} shows position-velocity maps of \hcop, \hcn\ and \methato, along a cut aligned with the two peaks of the \methato, and passing through the continuum peak of Per-bolo 45 (cyan dashed line in Figure~\ref{fig:perbol45_moments}). A red-shifted feature, about 1.4 \kms\ away from the cloud velocity (8.6 \kms), is observed in all lines except for \htcop. However, the lack of corresponding blue-shifted feature and the low-velocity and morphology of this red-shifted feature prevent us from concluding that this could be associated with an outflow from Per-bolo 45 rather than, for example infall. Consistent with this, no clear outflow is detected in the CO(2-1) observations of the MASSES survey \cite{2018StephensMass,2019StephensMass}. Their CO(2-1) map shows a redshifted feature to the south and west of the continuum source, surrounding the structure traced by \nhtd\ and \nthp. This emission was proposed by \cite{2013PlunkettCARMA} as the tip of a redshifted outflow lobe from the source SVS 13c, located 180" ($\sim$0.2 pc) to the South-West of Per-bolo 45. The interaction of this outflow with Per-bolo 45 dense core would provide a natural explanation to some of the higher-velocity emission seen in the maps of \hcop, \hcn\ and \methato, as well as to the SiO emission \citep{2010Jimenez-SerraParsec} that was observed in \cite{2012SchneeHow}.\\

The  properties of the continuum emission and the dense gas tracers in our sample, as well as the lack of conclusive evidence of an outflow strongly suggest that Per-bolo 45 is prestellar in nature, not having formed a compact object such a FHSC yet. 

\subsection{Per-bolo 58: a young Class 0 protostar}
\label{sec:per58_analysis}
\begin{figure*}
\includegraphics[width=1\textwidth]{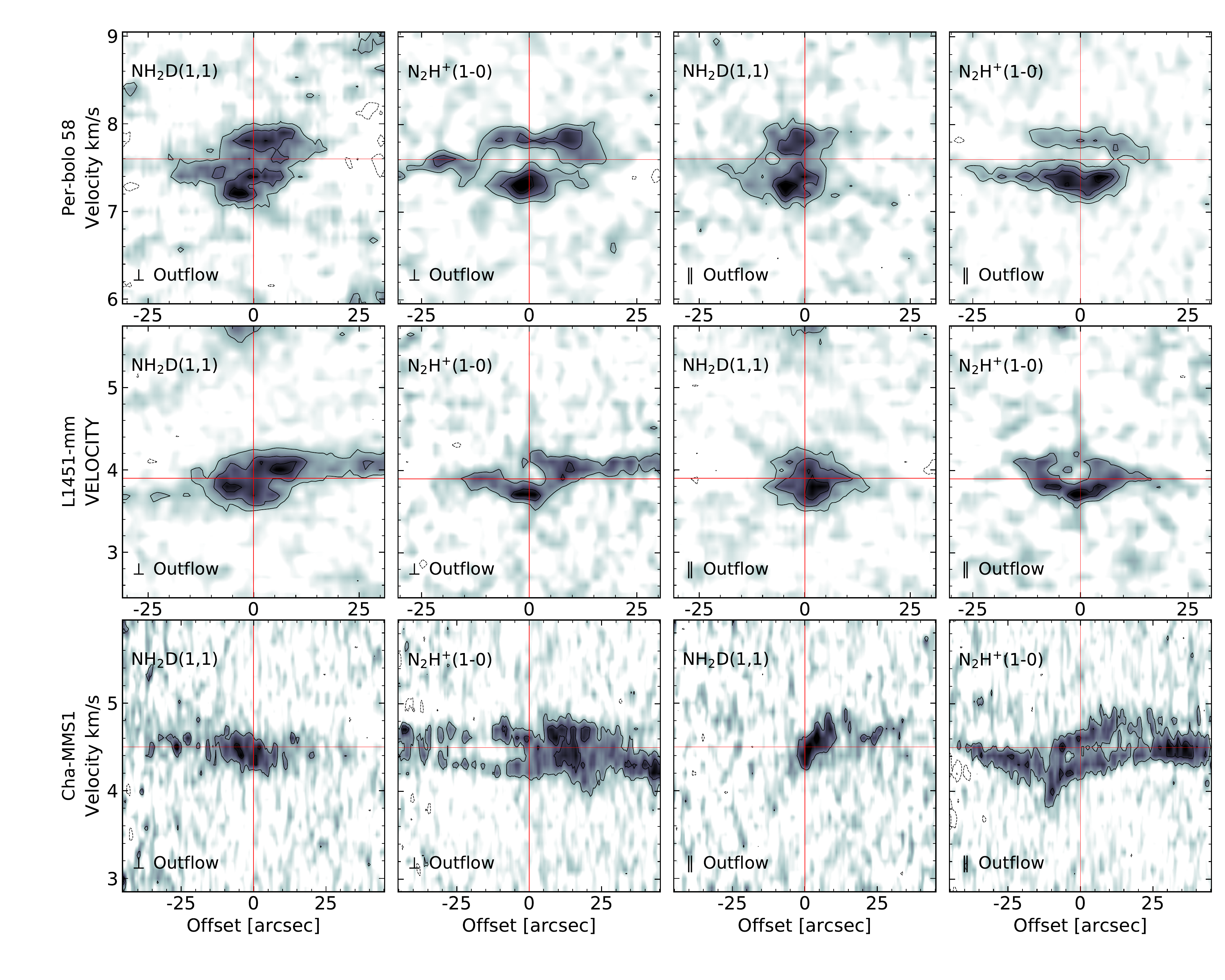}
    \caption{Per-bolo 58 (top), L1451-mm (middle) and Cha-MMS1 (bottom) position-velocity diagrams for \nhtd\ and \nthp, at cuts perpendicular and parallel to the outflow. 
    For Per-bolo 58 and Cha-MMS1, the most blue-shifted hyperfine of \nhtd\ is shown. For L1451-mm, individual satellites were weak. Hence, we show the average of the four satellites (see \ref{ap:hyperfines}). The systemic velocity is marked with a horizontal line at $v=7.6$\kms, $v=3.9$\kms\ and $v=4.5$\kms\ for Per-bolo 58, L1451-mm and Cha-MMS1, respectively. Contours start at 3$\sigma$ and increase in steps of 3$\sigma$, with $\sigma$ listed in Table~\ref{tb:cubes_properties}) for each transition. The direction of the cuts is shown at the bottom-left corner of each panel. 
    \label{fig:pvmaps_nh2d_n2hp_outsources}}
\end{figure*}

Per-Bolo 58 is a dense Far-IR/sub-mm core located in the northern part of the NGC 1333 region in Perseus. It was identified as a FHSC candidate by \cite{2010EnochCandidate} based on weak detections at 24 $\mu$m and 70 $\mu$m in Spitzer observations. Follow-up SMA observations at a resolution of 2.7" ($\sim$780 au), revealed a 1.3 mm compact source and a CO(2-1) bipolar outflow \citep{2011DunhamDetection}. The outflow appears jet-like with a characteristic projected velocity of 2.9 \kms. If the source is close to edge-on, the jet-like outflow velocity may be significantly faster, contrary to the typical properties of the FHSC outflow in simulations (\citealt{2008MachidaHigh,2014BateCollapse}).

In \cite{2017MaureiraTurbulent} we studied this source with 3 mm CARMA observations of molecular lines, at a resolution of 6" (NH$_2$D, N$_2$H$^+$,HCN, HCO$^+$, CS). The CARMA observations revealed line profiles with two distinct peaks separated by 0.4-0.6~\kms, arising from two different optically thin velocity components along the line of sight. These two velocity components are also present in our ALMA observations, including in the optically thin line \htcop. The two velocity components spatially overlap in a region that is a few  $10^3$ au in size towards the center of the core, 
 seen as the region with significantly higher  values of the line width in the moment 2 maps (see Figure~\ref{fig:perbol58_moments}). 
In \cite{2017MaureiraTurbulent} we concluded that the detected profiles with two velocity components could arise from infall motions in an in-homogeneous and flattened envelope seen close to edge-on 
and/or as a consequence of accretion from large-scale filamentary structures. 
\\  

A unique future among our sample, Figure~\ref{fig:perbol58_moments} shows that the \htcop\ emission is bright close to the central region of the core. 
This provides independent evidence that the evolutionary stage of this source is more evolved that the FHSC phase. This is because the size of the region in which carbon-bearing molecules are "back in" the gas phase (after being released from the dust icy mantles due to heat from the central source) increases with time \citep{2004LeeEvolution,2005JorgensenMolecular}. In Per-bolo 58 this region is large enough to be  detectable in our $\sim1,000$ resolution observations.

\subsubsection{Infall and rotation}
\label{sec:per58_kine}

Figure~\ref{fig:pvmaps_nh2d_n2hp_outsources} (top panels) shows position-velocity diagrams of the most blue-shifted satellite of the \nhtd \/ line, and the isolated hyperfine component of \nthp. The p-v diagrams correspond to cuts perpendicular and along the outflow. Perpendicular to the outflow, the higher intensity contours of both molecules show a hint of a velocity gradient (from the bottom-left to the upper-right quadrant). Along the outflow direction, no clear gradient is observed. However, there is an indication of higher velocity towards the center in  \nthp. These signatures, a gradient perpendicular to the outflow and higher velocities towards the source location, are expected when there is a combination of rotational and infalling motions \citep{2001HogerheijdeInfall}.

\begin{figure*}
\includegraphics[width=1\textwidth]{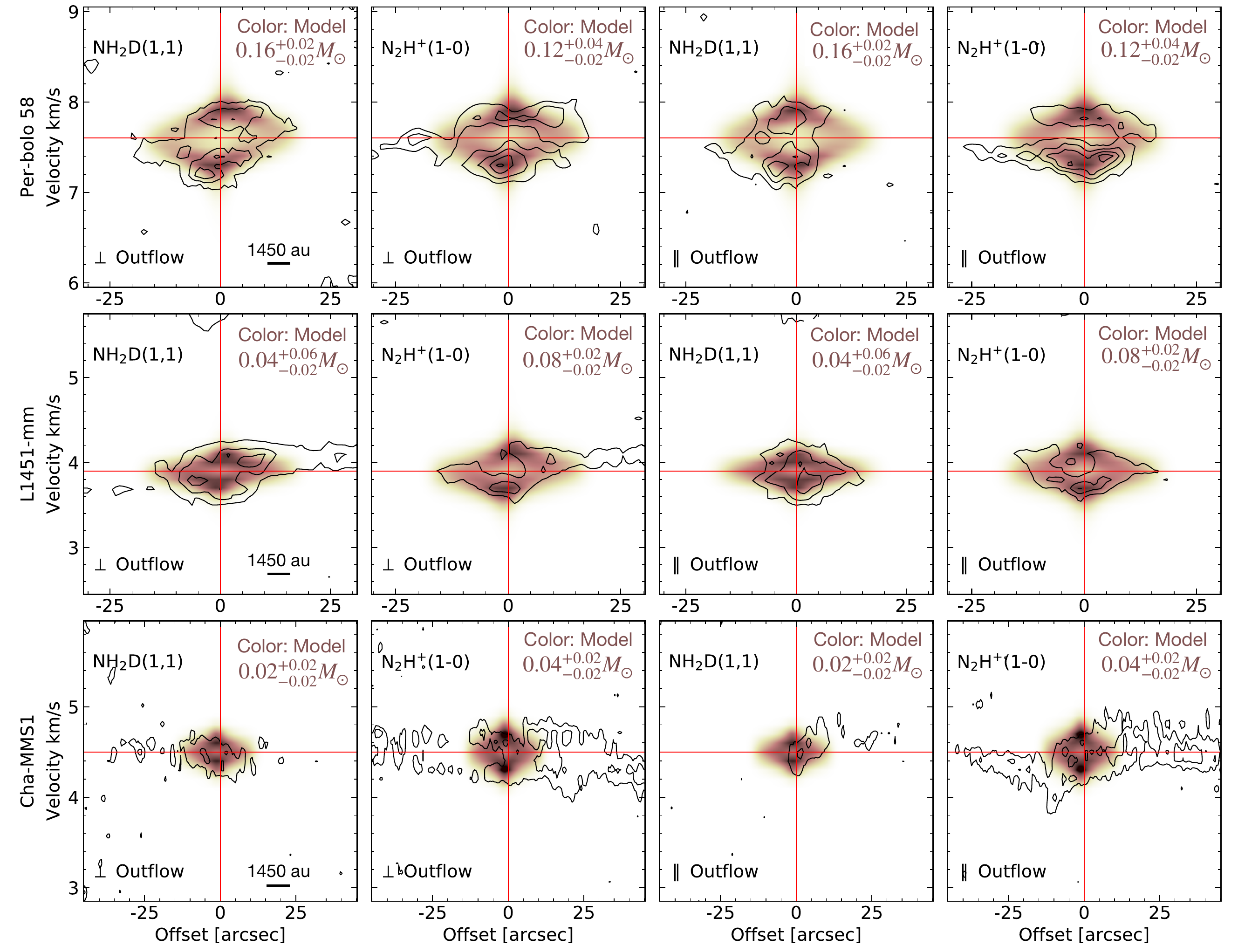}
\caption{Position-velocity diagrams of \nhtd\ and \nthp (shown in black contours) of Per-bolo 58 (top), L1451-mm (middle) and Cha-MMS1 (bottom)  at cuts perpendicular and parallel to the outflow of each source (same as Figure~\ref{fig:pvmaps_nh2d_n2hp_outsources}). The best infalling and rotating envelope model for each source and line is shown in color (brown-scale). The central mass inferred by the model is indicated at the upper right corner of each panel. \label{fig:pvmaps_mod}}
\end{figure*}

In order to estimate the mass of the central object that can produce the velocity structure observed in the p-v diagrams, we compare the position-velocity cuts with a toy model that assumes infall and rotation with conservation of angular momentum \citep{1976UlrichInfall}. Here we give a brief description of the model, for further details see appendix~\ref{ap:model}. The free parameters are the mass of the central object and the disk radius (which is assumed to be twice the radius of the centrifugal barrier). We do not expect to reproduce the emission at every position since clearly, the envelope structure is more complex than a spherical toy model. Likewise, a caveat of this model is that it assumes that the mass of the system is dominated by the central object mass. This might not be the case for extremely young objects such a FHSC, because such a young object might have proportionately more mass in the envelope compared to central region with respect to more evolved protostars. Given this, our estimated central mass can be taken as an upper limit. On the other hand, the model does not consider velocities smaller than those expected from free-fall (for instance in the presence of magnetic fields), which nevertheless have been observed at higher resolution towards some sources (e.g., \citealt{2014OhashiFormation}) and may result in an underestimation of the central object’s mass by a factor of a few \citep{2015YenObservations}. Despite these caveats, the model is still useful for comparing the FHSC candidates within the sample and Ulrich models applied to older objects including extremely young Class 0 protostars \citep{2013YenUnveiling,2014OyaSubstellar,2015YenObservations}. Models are calculated for a grid of masses and disk sizes and the best model for each source and species correspond to that with the least residual between the observed and modeled position-velocity maps. The best models for NH$_2$D and N$_2$H$^+$ are shown in Figure~\ref{fig:pvmaps_mod} in color. They indicate values for the mass of the central object of $0.16^{+0.02}_{-0.02}$ \msun\ and $0.12^{+0.04}_{-0.02}$, respectively. The disk radius of the best models range from 100 to 300 au. We note however that at the scales of our observations infall dominates over rotation and our constraint on the disk radius is poor. The value of the mass is slightly larger than what would be expected for a FHSC ($\approx$ 0.1 \msun, \citealt{2015TomidaRadiation}).

\begin{figure*}
\includegraphics[width=1\textwidth]{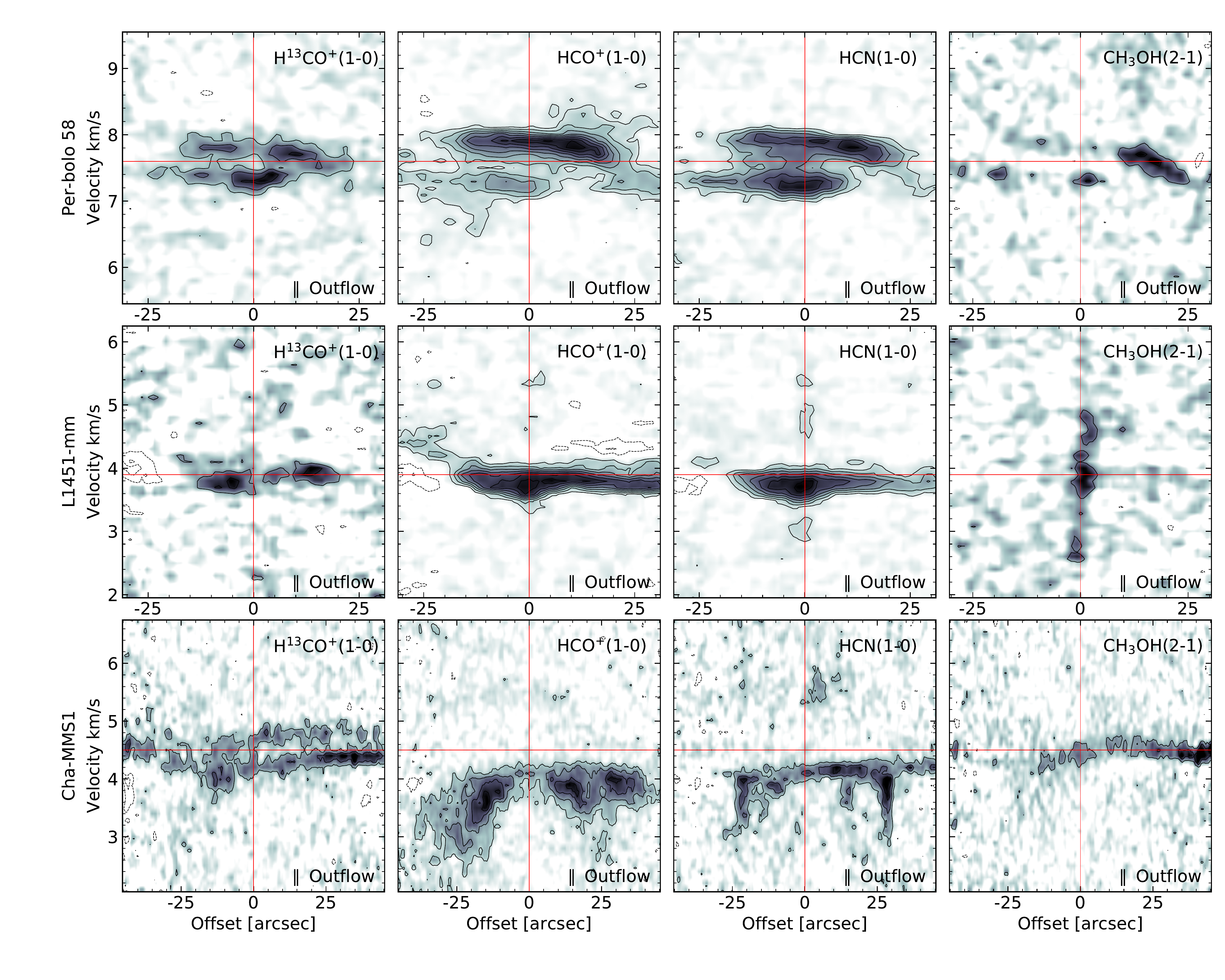}
    \caption{Per-bolo 58 (top), L1451-mm (middle) and Cha-MMS1 (bottom) position-velocity diagrams for \htcop, \hcop,\hcn\ and \methato\ at cuts along the outflow direction of each source. The zero offset corresponds to the position of the continuum source, listed in Table~\ref{tb:cont_fit} and the zero velocity corresponds to the systemic velocity of the sources ($7.6$\kms, $3.9$\kms\ and $4.5$\kms\ for Per-bolo 58, L1451-mm and Cha-MMS1, respectively). Contours start at 3$\sigma$ and increase in steps of 3$\sigma$, with $\sigma$ listed in  Table~\ref{tb:cubes_properties}.)
    \label{fig:pvmaps_alongoutflows}}
\end{figure*}

\subsubsection{Outflow}

The outflow is clearly identified only in the HCO$^+$  (Figure~\ref{fig:pvmaps_alongoutflows}). Emission at velocities 0.5-1\kms\ away from the source's systemic velocity is observed at distances $\gtrsim$ 3,000 au from the continuum location, consistent with the $\sim$7,000 au CO outflow lobe length observed by \citet{2011DunhamDetection} with SMA observations. The lobe length measured in the SMA observations should also be considered a lower limit for the outflow's true extent as outflow emission is detected right up to the edges of the primary beam.\\

In summary, the discussed presence of brighter \htcop\ emission towards the center of this source, the extension and properties of the CO outflow \citep{2011DunhamDetection} and our mass estimate for Per-bolo 58 are all consistent with this source more likely being a young (Class 0) protostar rather than a FHSC. 

\subsection{L1451-mm: a promising FHSC candidate}

L1451-mm is a dense Far-IR/sub-mm core in the quiescent dark cloud L1451 in the western end of the Perseus molecular cloud. This region appears to be in an early evolutionary state, with no Spitzer-identified YSOs at wavelengths shorter than 160 \micron. Observations with the SMA with a resolution of 1" ($\sim$290 au), revealed a 1.3 mm continuum point source and an unresolved, low-velocity ($\sim2$\kms) CO outflow. This is, to the best of our knowledge, the smallest (lobe length $\sim$500 au) and weakest outflow detected in a low-mass star-forming core (\citealt{2011PinedaEnigmatic,2011DunhamDetection}).\\

Among all our observed lines towards L1451-mm, H$^{13}$CO$^+$(1-0) is the only one that does not peak at the position of the continuum, showing a lack of emission within $\sim$1,000 au from it. The absence of emission from this molecule towards the center of the core suggests that carbon-bearing molecules are still frozen onto dust grains towards the center of L1451-mm, as typically seen in prestellar cores \citep{2002CaselliMolecular}. This is also consistent with the non-detection of C$^{18}$O (2-1) in SMA observations of this candidate, at a sensitivity of $\sim$0.3 K \citep{2019StephensMass}, further supporting an extremely young evolutionary state for this object. 
We note that HCO$^{+}$ and HCN are significantly more abundant than H$^{13}$CO$^{+}$ and can be found in the gas state in the outer layers of cores even when these carbon-bearing species are locked in the ice in the inner cold regions of the core. Thus, HCO$^{+}$(1-0) and HCN(1-0) emission from the outer layers of the core can contribute significantly to the total integrated intensity of the core map, resulting in peaks towards the center of the moment 0 maps of these two species, while the optically thinner \htcop\ shows a hole.

\subsubsection{Infall and rotation}
\label{sec:l1451mm_kin}

Similar to Per-bolo 58, the \nhtd\ and \nthp\ show velocity gradients perpendicular to the outflow and higher velocities towards the center (middle panel Figure~\ref{fig:pvmaps_nh2d_n2hp_outsources}). We interpret these observations as evidence for rotation and infall motions. We use the same model as for Per-Bolo 58 (see Appendix~\ref{ap:model}). The best model, shown in Figure~\ref{fig:pvmaps_mod}, results in a central mass of $0.04^{+0.06}_{-0.02}$ \msun\ and $0.08^{+0.02}_{-0.02}$ \msun\ using \nhtd\ and \nthp, respectively. The disk radius of the best models ranges from 100 to 300 au. (i.e, the full range of values we explored, except from a value  of 0 au, or no disk). The low mass estimate is consistent with both a FHSC or a very young protostar. To the best of our knowledge only three protostars to date show similar low masses when their envelopes are observed at similar resolution and a similar method is used to derive their mass. These are: IRAS 16253-2429: 0.03 \msun, NGC 1333 IRAS 4B: 0.07 \msun, L1157-mm: 0.05 \msun, from the \citealt{2015YenObservations} survey that studied 17 Class 0/I sources. Most sources in that survey have masses that are greater than $0.1$~\msun, similar to our estimate for Per-bolo 58. Furthermore, for these few protostellar sources with estimated masses smaller than 0.1~\msun, the C$^{18}$O has been clearly detected at $\sim$700-1,000 au scales \citep{2015YenObservations}, while for L1451-mm it remains undetected with a resolution of $\sim$1,000 au at comparable sensitivities \citep{2018StephensMass}. 

\subsubsection{Outflow}

Among the different lines we observed, the outflow in this source is only clearly detected in the \hcn\ and  \methato\ maps. In these lines a significantly larger velocity width is seen  in the moment 2 maps in a region  close to the continuum peak,
with an approximately north-south morphology
(see Figure~\ref{fig:l1451mm_moments}). Evidence of the existence of a very compact outflow is also seen in the p-v diagrams along the outflow axis, where red- and blue-shifted emission with velocities of    
 up to $\pm$1.5\kms\ away from the systemic velocity is seen  the \hcn\ and  \methato\  lines close to the continuum peak position (Figure~\ref{fig:pvmaps_alongoutflows}).
 Hints of  a red-shifted outflow lobe are also seen in the p-v of the \hcop\ line. In all cases, the high-velocity emission region is restricted to within a beam from the continuum peak position ($\lesssim10^{3}$ au). The size, the approximately north-south morphology, as well as the velocities of the putative outflow emission in our observations are consistent with the CO outflow in L1451-mm observed by   \cite{2011PinedaEnigmatic}.\\

The properties observed towards L1451-mm are extremely rare among typical Class 0 sources, and suggest that it is one of the youngest (if not the youngest) known low-mass protostellar source and thus  a promising FHSC candidate. Higher-resolution observations are required to further constrain the true evolutionary state of this source (discussed in  Section~\ref{sec:discussion}).

\subsection{Cha-MMS1: a young star-forming core interacting with a nearby outflow}

\begin{figure*}
\includegraphics[width=1\textwidth]{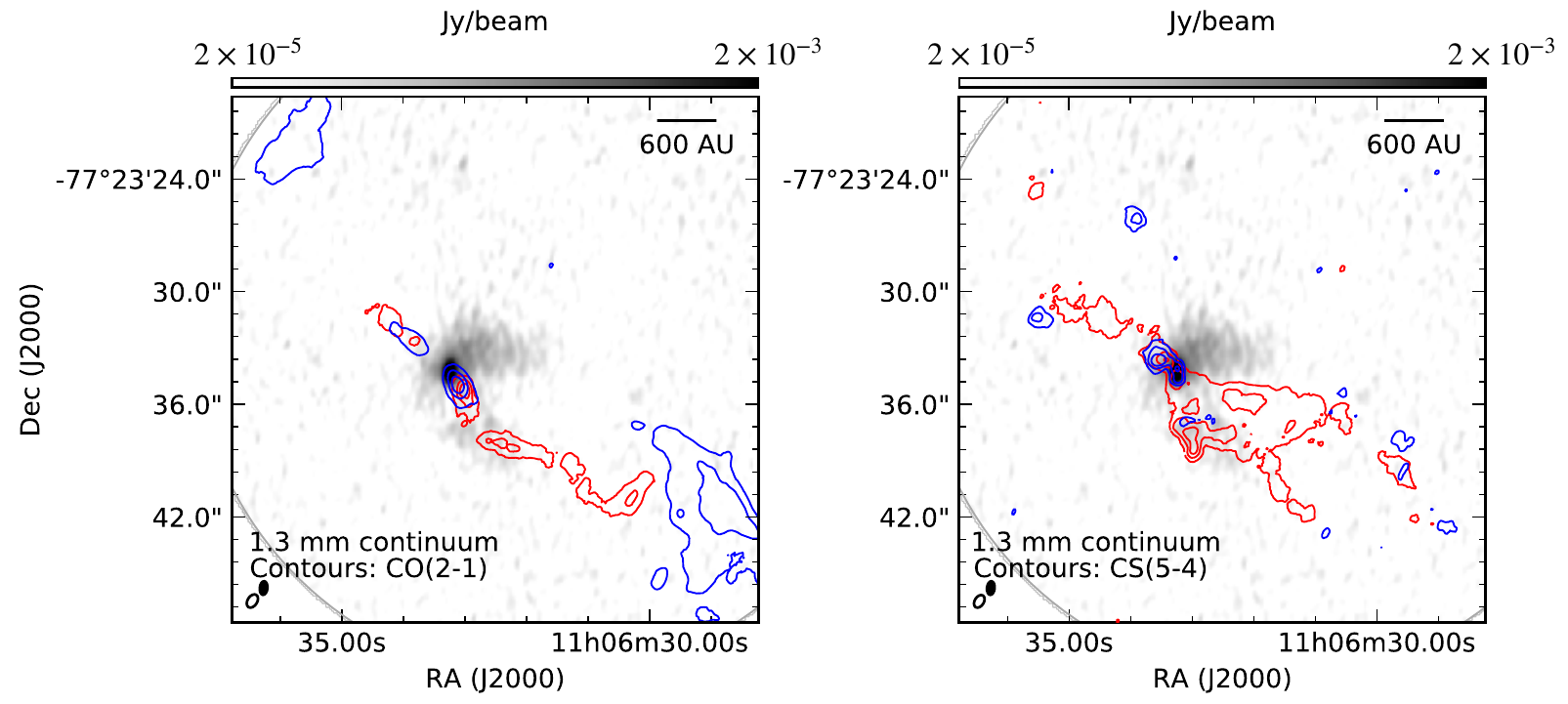}
    \caption{ALMA observations showing the outflow driven by Cha-MMS1. Both panels show the 1.3 mm continuum emission in grey-scale. {\it Left:} Contours show CO(2-1) blue-shifted (1.9-3.6~\kms) and redshifted (5.8-14.7~\kms) integrated intensity (the systemic velocity of this source is 4.5~\kms).
    Contours start at 10\% of the maximum and increase in steps of 30\%. The maxima of the blue-shifted and red-shifted emission are 0.2 and 0.1 Jy beam$^{-1}$~\kms, respectively. 
    {\it Right:} Contours show CS(5-4) blue-shifted (4.0-4.2~\kms) and redshifted (4.6-5.5~\kms) integrated intensity. The maxima of the blue-shifted and red-shifted emission are 0.01 and 0.04 Jy beam$^{-1}$~\kms, respectively. The beam sizes are indicated in the bottom left corner with a filled and open ellipse for the continuum and molecular line, respectively. The solid grey line marks the 20\% level of the primary beam. 
    \label{fig:cham_co_outflow}}
\end{figure*}

Cha-MMS1 is a dense core in the Chamaeleon I molecular cloud, in a region known as Ced 110. \cite{2006BellocheEvolutionary} proposed it as a FHSC candidates based on weak emission at 24 $\mu$m and 70 $\mu$m with Spitzer. They did not detect any outflow from this source using CO(3-2) APEX observations (beam of $\sim$3,000 au). NH$_3$ observations by \citet{2014VaisalaHigh} with a resolution of $\sim$ 1,300 au 
report only tentative evidence of an outflow. Recently, \cite{2019BuschDynamically} showed CO(3-2) ALMA observations with a resolution of $\sim$0.5" (100 au) that confirm Cha-MMS1 as the driving source of a small outflow, with a project lobe lenght of about 2,400 au. They determined deprojected outflow speeds of tens of~\kms, in disagreement with model predictions for outflows driven by FHSCs. In Section~\ref{sec:cham_outflow} we discuss complementary Archive ALMA cycle 2 observations of CO(2-1) and CS(5-4) towards the outflow (Figure~\ref{fig:cham_co_outflow}). \\

\begin{figure*}
\includegraphics[width=0.9\textwidth]{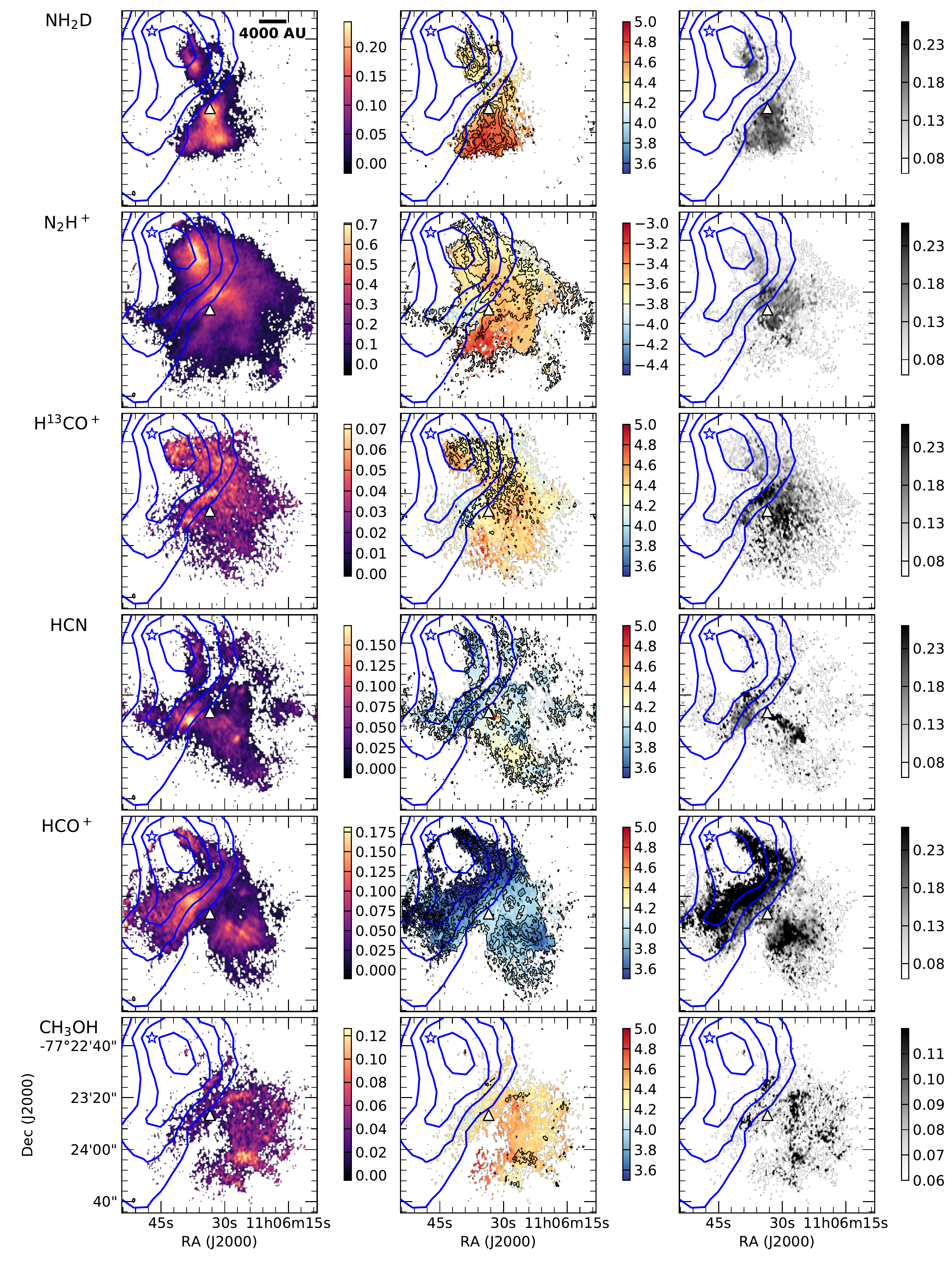} 
    \caption{Cha-MMS1, same as Figure~\ref{fig:cham_moments} but with the addition of the ALMA Total Power Array observations. Blue contours show the CO(3-2) integrated intensity for the velocity range between 0.4 and 3 \kms, reproduced from \protect\cite{2011LaddInteractions} and \protect\cite{2007HiramatsuAste}. 
    The position of IRS4 (upper left corner) is marked with a star.
    \label{fig:cham_co32_ladd}}
\end{figure*}

Cha-MMS1 is one of nine young low-mass stars within a radius of $\sim 0.2$ pc. The outflow from the nearby Class I source IRS4 (14,000 au to the north-east of Cha-MMS1) appears to be interacting with Cha-MMS1 envelope (\citealt{2006BellocheEvolutionary,2007HiramatsuAste,2011LaddInteractions}). Our ALMA maps also show evidence of this interaction. Figure~\ref{fig:cham_co32_ladd} shows the single-dish CO(3-2) blue-shifted (0-3 \kms) outflow emission arising from IRS 4 (from \citealt{2007HiramatsuAste}) overlaid on our ALMA (12m, 7m and TP combined) molecular lines maps. The moments maps show that the IRS4 outflow appears to change direction at the position of  high intensity emission of our high-density tracer (NH$_2$D, see Figure~\ref{fig:cham_co32_ladd}). In addition, the \nthp, \htcop\ and \methato\ show blue-shifted velocities and/or broad linewidths at the south-western edge of the IRS4 outflow.  In the \hcop\ and \hcn\ maps we also see enhanced emission in the regions where the south-western edge of the IRS 4 outflow  overlaps with the Cha-MMS1 core. This is similar to what we see in the L1448 region (Figure~\ref{fig:irs2e_moments}), and it is likely related to a  shock-induced abundance enhancement of these molecules. Our observations therefore provide further support to the scenario that the IRS4 outflow is interacting with (and it is deflected by) the Cha-MMS1 dense envelope material.\\

In our \nhtd\ observations, the peak emission is offset by approximately 850 au South-West from Cha-MMS1 (Figure~\ref{fig:cham_moments}). It is not clear if this is due to the source's evolution or because the outflow from IRS 4 is pushing the dense gas towards the South-West. The steep gradient in intensity of  \nhtd\ at the border facing the IRS 4 outflow, as well as the velocity gradient (in the same direction) suggest that the \nhtd\ could be offset due to the impact from the outflow. Similarly, the \nthp, although present at the position of the source, is weak in comparison with other positions in the region, and does not peak at the continuum peak position, consistent with the source being firmly in the Class 0/I phase. However, the \htcop, \hcn, \hcop\ and \methato\ are all weak and/or not present towards the compact continuum source (Figures~\ref{fig:cham_moments} and~\ref{fig:cham_spec}), as expected for sources in an extremely young evolutionary state, FHSC or very young Class 0 protostar.

\subsubsection{Infall and rotation}

Figure~\ref{fig:pvmaps_nh2d_n2hp_outsources} shows the position-velocity diagrams of the \nhtd\ and \nthp\ along a cut perpendicular and parallel to the outflow of this source (PA$_{out}=46^{\circ}$, Figure~\ref{fig:cham_co_outflow}). Along the outflow direction, both lines show a velocity gradient, possibly related to the interaction with the outflow powered by IRS4. Perpendicular to the outflow, both lines show almost no velocity gradient, which can be interpreted as small or no rotational motions at these scales (500 au to few 1,000 au). The \nthp\ shows a double peak profile in both cuts. These velocity components correspond to the blue-shifted structure to the North-East and red-shifted structure to the South-West of the source seen in the moment maps (Figure~\ref{fig:cham_moments}). We use our kinematic model (appendix~\ref{ap:model}) to provide an upper limit to the mass of the central object. We obtain values of $0.02^{+0.02}_{-0.02}$ \msun\ and $0.04^{+0.02}_{-0.02}$ \msun\ using \nhtd\ and \nthp, respectively. The estimate of the disk radius from our best models range from 0 to 300 au. Unlike the results for Per-bolo 58 and L1451-mm that required a disk of at least 100 au in radius. Similar to L1451-mm, the estimated upper limit to the mass is small, compared with observations of Class 0 and 0/I protostars at comparable scales and using similar methods \citep{2015YenObservations}.

\subsubsection{Outflow}
\label{sec:cham_outflow}

ALMA archive observations of CO (2-1) with a resolution of $\sim$0.5" (100 au) show the outflow driven by Cha-MMS1. Figure~\ref{fig:cham_co_outflow} exhibit blue- and red-shifted 
outflow velocities  
between about 1 to 3\kms and between about 1 and 10 \kms, respectively. 
In \cite{2019BuschDynamically}, they detect CO (3-2) blueshifted outflow velocities of up to about  11~\kms, and redshifted outflow velocities of up to 17~\kms, thus tracing higher-velocity gas than the CO (2-1) observations (which could due to the lower sensitivity, by a factor of two, in observations of the lower transition). The CO outflow shows mostly a jet-like morphology with a P.A. of $\sim$46$^{\circ}$. The blue-shifted and red-shifted emission overlap on both sides of the continuum emission, indicating an orientation close to the plane-of-sky. \citet{2019BuschDynamically} derive an angle of less than 20 degrees between the outflow axis and the plane-of-sky. The maximum projected extension as measured in the South-West lobe is about 2,700 au. The estimate of the projected dynamical time is $\lesssim 10,000$ years, similar to other FHSC candidates \citep{2011DunhamDetection}. 

Figure~\ref{fig:cham_co_outflow} also shows the integrated redshifted and blueshifted emission from CS(5-4). The velocity ranges of the integrated intensity are 4.0-4.2~\kms and 4.6-5.5~\kms, and the systemic velocity of the source is 4.5~\kms. Similar to the CO outflow, the  CS outflow  show  blue- and red-shifted emission in both lobes. However,  the morphology of the South-West CS lobe shows a clear cone-shaped structure, unlike the jet-like structure traced by the CO.
Our Cycle 1 ALMA \hcn\ observations also evidence of outflow emission. This can be seen in Figures~\ref{fig:cham_moments} and~\ref{fig:pvmaps_alongoutflows} which show  red-shifted emission with velocities about 1~\kms\ away from the systemic velocity, and consistent with the direction of the CO and CS outflow. A detailed characterization of the morphology of the CO outflow is beyond the scope of this paper, and will be addressed in a follow-up paper.

Summarizing, our ALMA maps and the small CO outflow are consistent with a very young evolutionary state for this candidate. Although we cannot securely discriminate between a FHSC or a young Class 0 protostar with the current data, the CO outflow morphology and a maximum projected outflow velocities of about 10~\kms\ suggests Cha-MMS1 already formed a protostar as suggested in previous independent studies (\citealt{2018YoungWhat,2019BuschDynamically}).

\section{Discussion}
\label{sec:discussion}

\begin{figure*}
\includegraphics[width=1\textwidth]{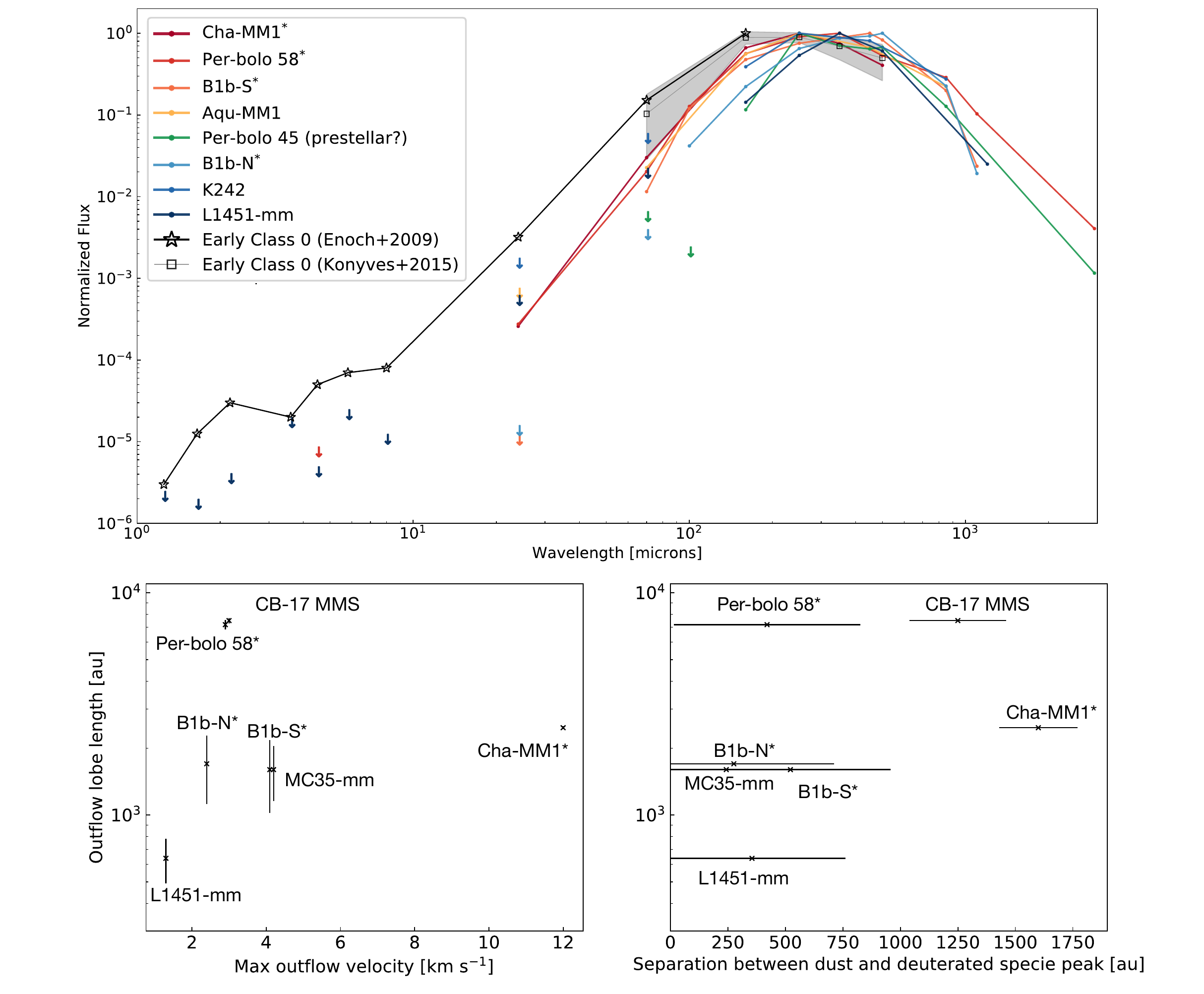}
    \caption{Comparison of current and former FHSC candidates in this work and the literature. Former FHSC candidates that have evidence for being young Class 0 sources are marked with $\star$ next to their names. 
    {\it Top:} Normalized spectral energy distributions. SEDs are normalized such that their peak correspond to a value of 1. Normalized SEDs for proposed FHSC candidates are shown in different colors. The average SED of early Class 0 sources (T$_{bol}\lesssim50$ K) from \citealt{2009EnochProperties} (black line with stars symbols) and the average SED of early Class 0 (T$_{bol}\lesssim50$ K) from \citealt{2011MauryFormation,2015KonyvesCensus} (gray line with square symbols) are also shown for comparison. The dispersion around the average for the later is shown as a gray shaded area. 
    {\it Bottom left:} CO molecular outflow lobe length vs. their maximum projected velocity. Error bars in the outflow lobe length correspond to half the beam minor axis of the observations of the respective sources. For Per-bolo 58, the outflow lobe length is an upper limit, since the outflow is detected out to the edge of the primary beam \citep{2011DunhamDetection}. 
    {\it Bottom right:} CO molecular outflow lobe length vs. the distance between the peak of the integrated emission of a deuterated specie (typically NH$_2$D or N$_2$D$^{+}$). The error bars for the outflow lobe length are the same as in the left panel and were omitted for clarity. The error bars for the distance between the peak of the integrated emission of a deuterated specie correspond to half the beam minor axis of the observations of the respective sources. Values in these figures and their references are listed in Table~\ref{tb:cand_summary}.} 
    \label{fig:candidates_comparison}
\end{figure*}

In Figure~\ref{fig:candidates_comparison} we attempt to summarize and compare the properties of former and current FHSC candidates presented in this work and the literature. In the top panel we show normalized SEDs for the sources proposed as FHSC candidates and compared them to young Class 0 protostars (T$_{bol}\lesssim50$ K) from the literature. 
From this comparison it is clear
that the sources proposed as FHSC candidates appear generally younger than the average SED for a young Class 0 protostar, as 
the SED of FHSC candidates peaks at longer wavelengths and all show lower normalized flux at 70 $\mu$m, compared to  the average Class 0 SEDs. Nevertheless, some FHSC candidates do look more evolved (Per-bolo 58, Cha-MMS1, B1b-S and Aqu-MM1) than others, as their 70 $\mu$m fluxes are close to the lower end of the range of the young Class 0 sources observed in Figure~\ref{fig:candidates_comparison}. These FHSC candidates with more evolved SED's (except Aqu-MM1\footnote{To date, there are no interferometric observations probing the existence of an outflow or the emission properties of dense gas tracers.}) show evidence in support of a young Class 0 stage, when viewed with interferometric observations. Most of the evidence comes from the detection of high-velocity components in their outflow (Cha-MMS1, \citealt{2019BuschDynamically}; B1b-S, \citealt{2019HiranoTwo}) and/or from the distribution of molecular line emission at the inner envelope or disk scales (Per-bolo 58, Section~\ref{sec:per58_analysis} in this work; B1b-S \citealt{2018MarcelinoALMA}). 

The rest of the FHSC candidates show normalized SEDs with significantly less flux at short wavelengths from the peak (compared to average Class 0 SEDs), consistent with them being at a younger evolutionary state. Among these, there is Per-bolo 45 for which our observations indicate it is likely prestellar (Section~\ref{sec:per_45}, see also \citealt{2019StephensMass}). Included in this group is also B1b-N which was recently confirmed as a young Class 0 protostar owing to the  newly detected high-velocity outflow component (projected maximum velocity of 9~\kms) traced by methanol \citep{2019HiranoTwo}. The two remaining sources, K242 and L1451-mm show the least evolved SEDs. The former was proposed as a FHSC by \cite{2018YoungWhat} based on modeling of the SED. However, there are no more constraints of the evolutionary state of this source as it currently lacks  interferometric observations. As for L1451-mm, in Section~\ref{sec:analysis_per_source}, we analyzed this source and concluded that it was consistent with either an extremely young Class 0 protostar or a FHSC. The bottom panels in Figure~\ref{fig:candidates_comparison} show a comparison of L1451-mm properties observed with interferometric observations with that of other FHSC candidates with outflow detection. In particular, the bottom left panel shows the CO outflow properties (maximum velocity, maximum lobe length) and the bottom right panel compares the distance between the dust emission peak and the emission peak of a cold/high-density gas tracer (such as NH$_2$D, N$_2$D$^+$). In both panels, the younger the source, the closer it should lie to the lower left corner. From these plots, the most promising FHSC candidate in our sample, L1451-mm, appears as the youngest source. Although more evolved than L1451-mm (according to its position in the bottom panels of Figure~\ref{fig:candidates_comparison}), the recently identified source in Taurus (MC35-mm, \citealt{2020FujishiroLow}) may still be considered  a FHSC candidate. Two other FHSC candidates can be found in the literature but are not shown in Figure ~\ref{fig:candidates_comparison} due to their lack of confirmed outflow emission. These are SM1N and Oph A-N6, both located in Ophiuchus. Although typically considered as prestellar in nature, recent ALMA high-resolution observations revealed that they are not resolved-out and instead remain detected and showing concentrated continuum emission at 100 au scales \citep{2018FriesenALMA}. The observations in \cite{2018FriesenALMA} show no clear evidence in the  CO (2-1) emission of outflows towards SM1N and Oph A-N6. Red-shifted emission on one side of Oph A-N6 extending up to $\sim$1,400 au is suggestive, but further observations perhaps with a different tracer are needed to confirm this, since as pointed out in \cite{2018FriesenALMA}, any outflow from this source will be difficult to detect due to contamination from VLA 16293's outflow. Both sources also show significant emission of deuterated species \citep{2012BourkeInitial,2014FriesenH2D+}, peaking at similar distances from the continuum as L1451-mm (see Table~\ref{tb:cand_summary}). \\

For these four remaining FHSC candidates (L1451-mm, MC35-mm, SM1N and Oph A-N6) which have been observed  at intermediate scales (few 100 au to few 1,000 au) a final confirmation of their true evolutionary state requires higher-resolution observations. For L1451-mm, the compact outflow needs to be resolved to investigate its morphology and kinematics, as a higher-velocity component (an indication of protostellar nature) could be revealed by observations with a beam smaller than 100 au, similar to the case of B1b-N \citep{2019HiranoTwo}. An additional goal of high-resolution observations for L1451-mm and the remaining youngest candidates should be to investigate the temperature and density profiles of the envelope at scales from few au to 100 au. This is because simulations show that the temperature remains lower than $\sim$ 30 K even at several tens of au up to 100 au from the center \citep{2014BateCollapse,2015TomidaRadiation,2016HincelinChemical,2019YoungSynthetic} during the FHSC stage. On the other hand, Class 0 sources show temperatures of 20-30 K or higher at scales of several 100 au (sufficient for thermal evaporation of CO) which results in the inner envelope and disk being easily detected using C$^{18}$O observations \citep{2015YenObservations,2017YenSigns,2018StephensMass}. This holds even in very low luminosity objects, for which the unexpected large extent of C$^{18}$O is interpreted as evidence of a previous burst of accretion \citep{2017FrimannProtostellar,2018HsiehProbing}. As for the density profile, simulations of the FHSC stage show a flat inner region, corresponding to the FHSC structure and extending up to $\sim$10 au \citep{2014BateCollapse,2013TomidaRadiation}. For a protostar, on the other hand, the density profile should  increase towards the central $\sim$1 au region \citep{2019YoungSynthetic}. Observations of the continuum emission with a resolution better than a few tens of au are likely required to model the emission and provide  a density and temperature profile that can probe the relevant scales. Additional line observations with a similar resolution can also help to further distinguish between the different models. We note that, as pointed out in \cite{2019YoungSynthetic}, distinguishing a dense core with only a FHSC and one that has recently formed a protostar but in which the FHSC structure is still present is likely not possible, even with high resolution observations. Given the optically thick nature of the FHSC core, it is difficult to probe the physical properties within the FHSC structure. Despite this, finding a source with density and temperature profiles as well as with outflow properties consistent with the theoretical predictions will provide convincing evidence in support of a bona fide FHSC.\\

Under the assumption of constant star formation rate, \cite{2011PinedaEnigmatic} estimated the  number of FHSC expected in the Perseus molecular cloud, as a fraction (proportional to the lifetime of a FHSC) of the current Class 0 sources. They concluded that the  number of candidates (Per-bolo 58, L1451-mm, Per-bolo 45 and L1448 IRS2E) was inconsistent with the theoretical predictions. We can reassess this analysis in light of our  results.
Using a lifetime of the Class 0 stage of 0.2 Myr \citep{2014DunhamEvolution}, then in the  scenario in which the FHSC lifetime is close to $10^4$ years, we would require at least 20 Class 0 sources to expect one FHSC. This number increases up to 200 if the lifetime is $10^3$ years. Since there are about 38 Class 0 sources in Perseus \citep{2015DunhamYoung}, one or two FHSC are expected if the lifetime is several $10^3$ years or more. This is consistent with our results which indicate there is only one FHSC candidate in Perseus  (L1451-mm).
 Given the low number of embedded protostars in the Chamaeleon I molecular cloud (< 10, \citealt{2016DunhamALMA}), the confirmation of Cha-MMS1 as the only Class 0 protostar (and not a FHSC) in this cloud is also in agreement with the theoretical expectations. Finally, we note that most of the proposed FHSC candidates have been found serendipitously. Unbiased interferometric surveys towards a large number of starless cores (e.g., \citealt{2016DunhamALMA,2017KirkALMA} and more recently \citealt{2020TokudaFragmentation}), are an efficient way to find sources on the verge of star formation, such a FHSC. Thus, future similar surveys towards regions with a larger number of starless and young protostellar sources are needed to improve the chances of observing a bonafide FHSC.

\section{Summary and conclusion}

We studied the emission properties and kinematics at envelope scales of five FHSC candidates; four located in the Perseus molecular cloud and one located in the Chamaeleon molecular cloud. We used 3 mm molecular lines and continuum ALMA observations with a resolution of 3.5"-4" ($\sim$700-1160 au). Our results can be summarized as follows.

\begin{enumerate}

\item
Two out of the five candidates remain consistent with a very young evolutionary state: L1451-mm and Cha-MMS1. The isolated source L1451-mm shows envelope and outflow properties that are extremely rare among typical Class 0 sources and even other FHSC candidates, which suggest that it is one of the youngest (if not the youngest) known low-mass protostellar source and thus  a promising FHSC candidate. Cha-MMS1's envelope does not show definitive evidence to rule it out as a FHSC candidate (or confirm it as a Class 0 protostar). Yet, the collimated morphology of its CO molecular outflow and high-deprojected velocities suggests Cha-MMS1 is already in the protostellar phase.  

\item
Among the remaining three sources in our sample (Per-bolo 58, Per-bolo 45 and L1448 IRS2E), we classify 
Per-bolo 58 as a young Class 0 protostar 
on the basis of 
the properties of its envelope and outflow, which are  inconsistent 
with those expected for FHSC. Per-bolo 45 continuum emission is elongated and almost resolved out in the interferometer data. This, along with a lack of conclusive evidence for an embedded outflow, firmly favors a prestellar (pre-FSHC) stage for this candidate. In the case of L1448 IRS2E, our observations support that there is no core (pre or protostellar) at this location (see below). 

\item
We did not detect any 3 mm continuum or evidence of dense gas (traced by N$_2$H$^{+}$ and NH$_2$D) towards the FHSC candidate L1448 IRS2E. Our maps show evidence that the gas in this location is impacted by the nearby outflow from the Class 0 protostellar system L1448 IRS2. Given the morphology of the carved gas as well as the velocities of the proposed L1448 IRS2E's CO outflow (and other outflow tracers), a more natural explanation for the non-detection of the continuum is that the red-shifted emission interpreted as an outflow from L1448 IRS2E by Chen et al. (2010) is emission from the nearby outflow that has been deflected by the dense gas present at the location where the FHSC candidate was thought to be. \\

\end{enumerate}

Our observations show that not all sources proposed as FHSC candidates, based on their SED and/or the detection of dust and outflow in interferometric observations, pass the test of truly being in an extremely young evolutionary state (FHSC or extremely young protostar). Thus, observations of the inner and outer envelope of FHSC candidates are a useful and efficient tool to identify the most promising candidates. The two youngest sources identified in this work (L1451-mm and Cha-MMS1) are rare and unique laboratories for investigating the earliest stages of star formation. Future observations at envelope-disk and disk scales down to 10 au resolution towards the most promising candidate in this work (L1451-mm) as well as those in the literature will serve to constrain models of core collapse and disk formation, as well as to further discriminate their true nature.

\section*{Acknowledgements}

MJM and HGA acknowledge support from the National Science Foundation award AST-1714710. D.M. acknowledges support from CONICYT project Basal AFB-170002. MJM and JEP are grateful for support from the Max Planck Society. \\

We thank Ned Ladd for providing us with the  CO(3-2) single dish data for Cha-MMS1. This paper makes use of the following ALMA data: ADS/JAO.ALMA\#2012.1.00394.S and ADS/JAO.ALMA\#2013.1.01113.S. ALMA is a partnership of ESO (representing its member states), NSF (USA) and NINS (Japan), together with NRC (Canada), MOST and ASIAA (Taiwan), and KASI (Republic of Korea), in cooperation with the Republic of Chile. The Joint ALMA Observatory is operated by ESO, AUI/NRAO and NAOJ.\\

{\bf DATA AVAILABILITY:} The raw data is available on the ALMA archive ADS/JAO.ALMA\#2012.1.00394.S, ADS/JAO.ALMA\#2013.1.01113.S. The calibrated and imaged data underlying this article will be shared on reasonable request to the corresponding author.


\bibliographystyle{mnras}
\bibliography{firstcors_alma}



\appendix

\section{Spectra at the position of the sources}
\label{ap:extrafigures}

\begin{figure}
\centering
\includegraphics[width=0.5\textwidth]{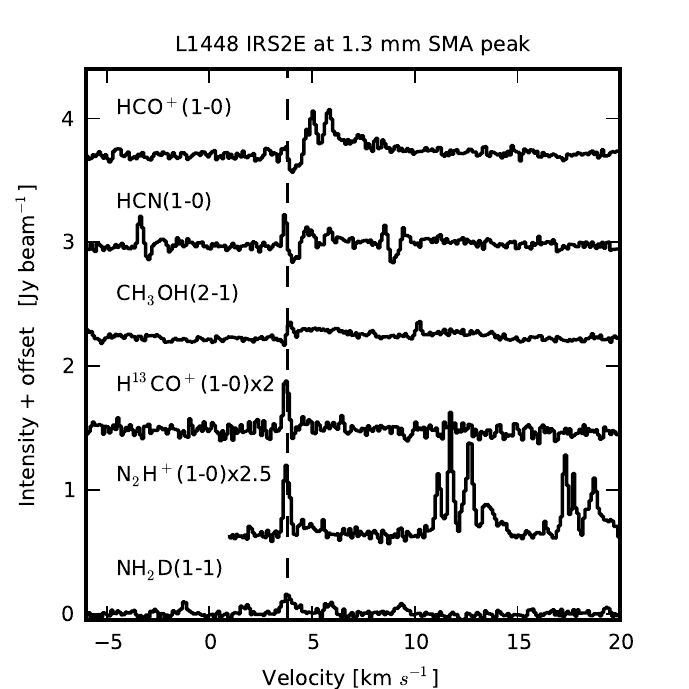}
    \caption{ L1448IRS2E molecular line spectra at the 1.3 mm SMA continuum peak in \protect\cite{2010ChenL1448}. The vertical lines corresponds to 3.8 km s$^{-1}$, the presumed systemic velocity of the envelope at the location of candidate. The spectra are averaged over a beam. \label{fig:irs2e_spec}}
\end{figure}

\begin{figure}
\centering
\includegraphics[width=0.5\textwidth]{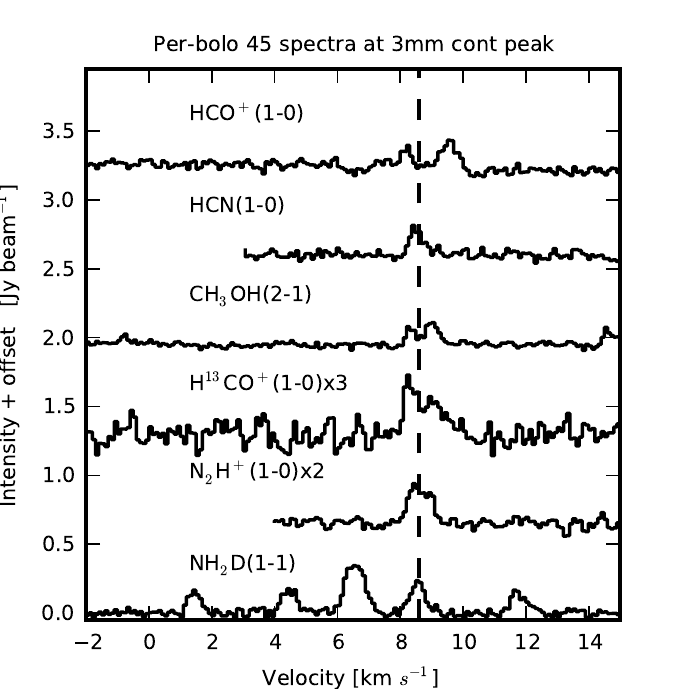}
    \caption{ Per-bolo 45 molecular line spectra averaged over a beam at the position of the 3mm continuum peak. The vertical line corresponds to 8.6 \kms, the presumed systemic velocity of the envelope at the location of candidate. The \nthp\ spectrum is shifted to show the isolated hyperfine component. The \nhtd\ and \hcn\ spectra are also shifted to show a satellite group of hyperfines lines and the weaker hyperfine line, respectively.
    \label{fig:per45_spec}}
\end{figure}

\begin{figure}
\centering
\includegraphics[width=0.5\textwidth]{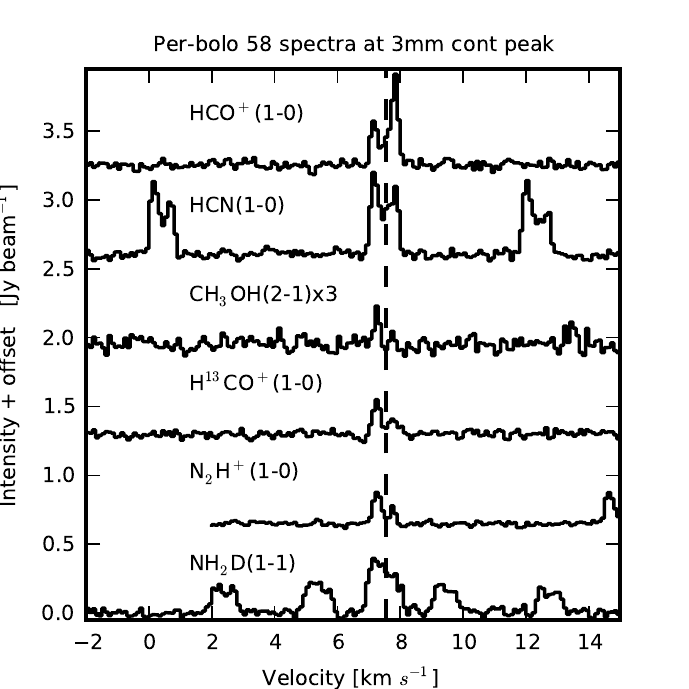}
    \caption{ Per-bolo 58 molecular line spectra averaged over a beam at the position of the 3mm continuum peak. The vertical line corresponds to $v=7.55$ \kms, the presumed systemic velocity of the envelope at the location of candidate. The \nthp\ spectrum is shifted to show the isolated hyperfine component.  \label{fig:per58_spec}}
\end{figure}

\begin{figure}
\centering
\includegraphics[width=0.5\textwidth]{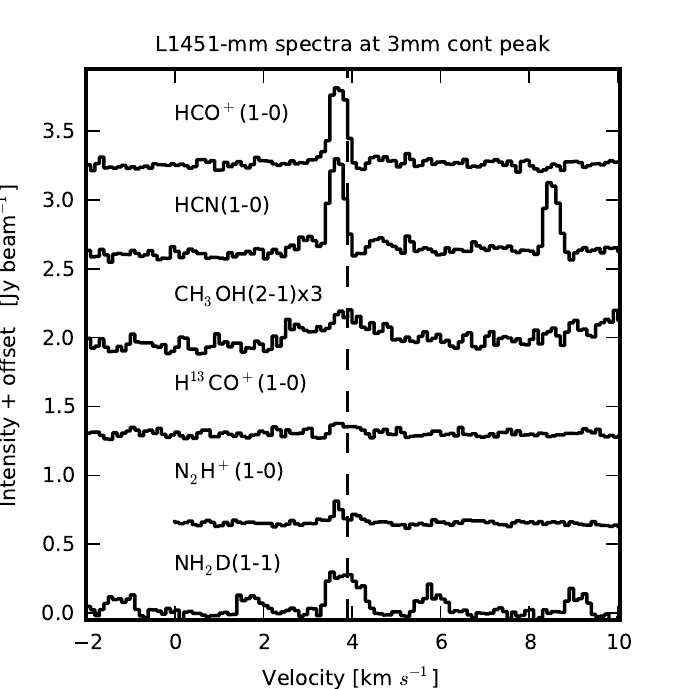}
    \caption{L1451-mm molecular line spectra averaged over a beam at the position of the 3mm continuum peak. The vertical line corresponds to $v=3.9$ \kms, the presumed systemic velocity of the envelope at the location of candidate. The \nthp\ spectrum is shifted to show the isolated hyperfine component.
    \label{fig:l1451mm_spec}}
\end{figure}

\begin{figure}
\centering
\includegraphics[width=0.5\textwidth]{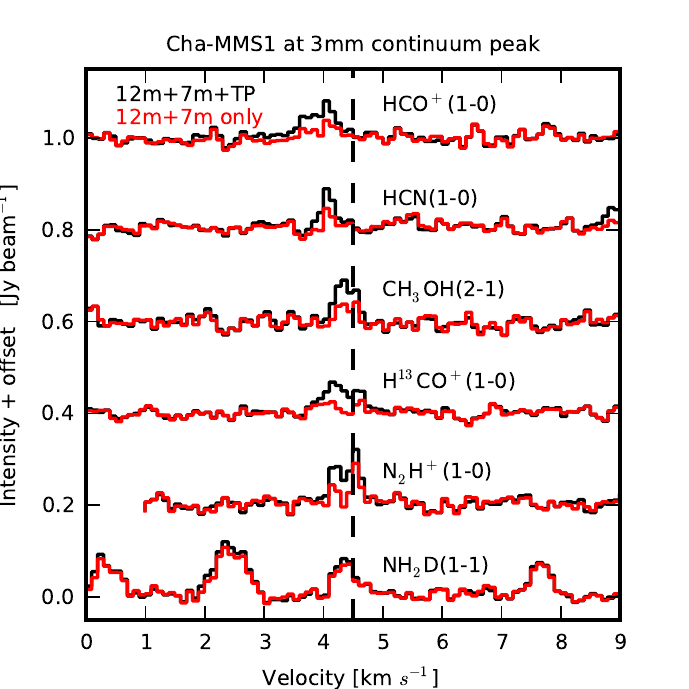}
    \caption{ Cha-MMS1 molecular line spectra averaged over a beam at the position of the 3mm continuum peak with (black) and without (red line) the TP observations. The vertical line corresponds to $v=4.5$ \kms, the presumed systemic velocity of the envelope at the location of candidate. The \nthp\ spectrum is shifted to show the isolated hyperfine component. The \nhtd\ spectrum is also shifted to show a satellite group of hyperfine lines.
    \label{fig:cham_spec}}
\end{figure}

Figures~\ref{fig:irs2e_spec} to~\ref{fig:cham_spec} show the spectra of all the species we observe towards the the position of the FHSC candidates in our study.

\section{Position-velocity map using the satellites of  NH\texorpdfstring{\textsubscript{2}}{2}D}

\label{ap:hyperfines}

For Per-bolo 58 and Cha-MMS1 the position-velocity maps in Figure~\ref{fig:pvmaps_nh2d_n2hp_outsources} were constructed using the most blue-shifted satellite. This satellite has 3 hyperfine components \citep{2016DanielNH2D} separated by less than our channel width (0.1 \kms). The middle hyperfine at 85.927724 GHz is the brightest and we use this frequency for the rest frequency when constructing the map. 

For L1451-mm the most blue-shifted satellite by itself was too faint to use it for kinematics analysis. Thus we constructed a position velocity map by averaging the position-velocity map of the four \nhtd\ satellites. Each satellite has one or more hyperfine lines with an intensity of at least 10\% of the brightest hyperfine within the satellite. These neighboring hyperfine lines are all within 0.2 \kms\ of the brightest of the group. Thus they are not resolved in our observations. The rest frame frequencies that we used for averaging the position-velocity maps are 85.927724 GHz, 85.926864 GHz, 85.925688 GHz, 85.924748 GHz, corresponding to the main hyperfine in each satellite.

\section{Infall and rotation toy model}
\label{ap:model}
\noindent The toy model considers the ballistic motion of particles under the influence of a central mass \citep{1976UlrichInfall}. The particles start from an spherical surface assumed to be in solid body rotation. The amount of rotation provides the total angular momentum which is conserved along the trajectories and results in a radius at which the infall velocity becomes zero and material cannot continue infalling (i.e., the centrifugal barrier). Before that, the infall velocity reaches its maximum value at a radius two times the centrifugal barrier value. This radius is typically assumed as the radius within which a rotationally supported disk can form \citep{1976UlrichInfall}. As described in \cite{1976UlrichInfall} the trajectories and velocities of the particles are fully defined by two parameters: the central mass and the disk (or centrifugal barrier) radius. All the trajectories in this model correspond to parabolas. 

As in \cite{2017MaureiraKinematics}, for a given pair of these parameters we construct a model cube to compare with observations. To produce the cube, given a coordinate position on the sky, we simulate a line of sight that crosses the spherical envelope at that position. The line of sight can have an inclination $i$ such such that if $i=90^{\circ}$, the observer is looking through the midplane in an edge-on configuration. We calculate the line of sight velocity every 10 au steps along a given line of sight. For each of these steps we calculate a Gaussian centered at that velocity with a linewidth equal to a thermal width at a temperature of 10 K. The amplitude of the Gaussian is set to be proportional to the density at that position assuming a density profile proportional to $r^{-2}$. The final spectrum along a line of sight is the sum of all the Gaussians along the line of sight. The cube was then convolved to match the beam of the observations and we construct position-velocity maps along the equator and perpendicular to it. To compare the kinematics between the observed and modeled data, for each spatial position along the p-v cut (i.e., for each column in the p-v diagrams) we matched the intensity peak value of the observed spectrum with the modeled one. For all sources we fix the inclination to  $i=90^{\circ}$ as this provides with an upper limit to the mass \citep{2017MaureiraKinematics}. Further details and discussion of the model are included in \cite{2017MaureiraKinematics}. 

For each source we constructed models with masses from 0.02 to 
0.22\msun\ in steps of 0.02\msun\ and centrifugal barrier radii of 0, 50, 100 and 150 au. Our models do not constrain well the radius of the centrifugal barrier, meaning that the appearance of the p-v diagrams do not change much for different values of this parameter in the range we explored. The upper limit of 150 au was chosen as a limiting extreme case, given that observations show that most protostellar disks are consistent with a radius smaller than 100 au (\citealt{2019MauryCharacterizing,2019PinedaSpecific,2018SeguraCoxVLA}). We note, however,  that increasing the centrifugal barrier radius results in larger values for the central mass that, in some cases, can match the data. For Per-bolo 58 and L1451-mm we set the radius of the initial sphere (envelope in our model) to 5000 au, while for Cha-MMS1 we fix it to 2700 au. These values were chosen as they match the extent of the emission observed in the p-v diagrams. 

We estimate the best-fit model and the uncertainty in the derived mass form the  model in the following way. For each p-v diagram and model, we calculated $res_{pix}$, in units of $\sigma$, corresponding to the residual (i.e., observation minus the model) divided by the rms of the p-v map. The total residual for each p-v diagram and model was then calculated as the maximum $res_{max}$ of the $res_{pix}$ values among all pixels within 15" of the source position. Then, for each model and molecule we calculated the combined residual $res_{comb}$ corresponding to the geometric mean of the $res_{max}$ values for the model p-v diagrams along the equator (perpendicular to the outflow in the observations) and perpendicular to the equator (parallel to the outflow in the observations). Finally, we report the best-fit models for each molecule as those that result in the minimum combined residual $res_{comb}$. The errors for the masses were calculated as those models that resulted in a $res_{comb}$ value up to 10\% higher than that of the best fit. This percentage usually corresponded to models that had a combined residual value that was approximately $1-sigma$ higher than the residual of the best model. We note that calculating residuals as the sum of $res_{pix}$ over all pixels in a p-v diagram results in similar 'best fit' values for the masses obtained using $res_{max}$ (within the provided uncertainties). However,  $res_{max}$ provided a more intuitive quantity to assess uncertainties in a uniform way for all sources.

\section{Properties of FHSC candidates from this work and the literature}
\label{ap:prop_disc}

Table~\ref{tb:cand_summary} lists the outflow and dense gas tracer properties used in Figure~\ref{fig:candidates_comparison} to compare current and former FHSC candidates in this work and the literature. 

\begin{table*}
\caption{Properties of current and former FHSC candidates on the literature}
\small
\label{tb:cand_summary}
\resizebox{\textwidth}{!}{\begin{tabular}{lccccc}

\hline
\hline
Source  &CO$_{out}$ lobe length$^{a}$& CO$_{out}$ velocity$^{a}$ & Distance to a deuterated specie peak$^b$ & Comments &References.\\
&[au]&[\kms]& [au]&&\\
\hline
\hline
L1448 IRS2 E &- &- & 8,600 (\nhtd) &ruled-out as a star forming core &1,5, 16 \\
Per-bolo 45 &- 	    &-	  &290$^{c}$ (\nhtd)	    &weak and extended continuum, likely prestellar &	1, 19\\
Per-bolo 58 &$>$7192&2.9  &420$^{c}$ (\nhtd, N$_2$D$^{+}$(3-2))	&young Class 0 protostar &	1,2,3,4,5\\
L1451-mm	&640	&1.3  &355$^{c}$ (\nhtd, N$_2$D$^{+}$(3-2))	&FHSC/young Class 0 protostar	 &1,5,6\\
Cha-MMS1	&2,470	&12	  &1600 (\nhtd)	    &young Class 0 protostar &	1,4,7,8,9\\
B1b-N	    &1,700	&2.4	  &276$^{c}$ (N$_2$D$^{+}$(3-2))	    &young Class 0 protostar &	5, 10, 11\\
B1b-S	    &1,600	&4.1	  &522$^{c}$ (N$_2$D$^{+}$(3-2))	    &young Class 0 protostar (hot corino) &	5,10, 11, 17\\
CB 17 MMS	&7,500	&3	  &1250 (N$_2$D$^{+}$(3-2))	    &FHSC/young Class 0 protostar	& 4, 12\\
SM1N	    &-	&-	  &100 (H$_2$D$^+$(1$_{1,0}$-1$_{1,1}$))	   &prestellar/FHSC/young Class 0 protostar	&13,14\\
Oph A-N6	&-	&-	  &700 (N$_2$D$^{+}$(3-2))	  &prestellar/FHSC/young Class 0 protostar&	13,15\\
Aqu-MM1 & - & - &- & prestellar/FHSC/young Class 0 & 4\\
K242 & - & - &- & prestellar/FHSC/young Class 0 & 4\\
MC35-mm & 1600 & 4.2 & 243$^{c,d}$ (N$_2$D$^{+}$(3-2)) & FHSC/young Class 0 protostar & 18\\
\hline
\hline
\end{tabular}}
\vspace{1ex}
\footnotesize
\raggedright REFERENCES: 
(1) This work, (2) \cite{2011DunhamDetection}, (3) \cite{2010EnochCandidate}, (4) \cite{2018YoungWhat}, (5) \citealt{2018StephensMass,2019StephensMass}, (6) \cite{2011PinedaEnigmatic}, (7) \cite{2019BuschDynamically}, (8) \cite{2013TsitaliDynamical}, (9) \cite{2014VaisalaHigh}, (10) \cite{2012PezzutoHerschel}, (11) \cite{2014HiranoTwo}, (12) \cite{2012ChenSubmillimeter}, (13) \cite{2018FriesenALMA}, (14) \cite{2014FriesenH2D+}, (15) \cite{2012BourkeInitial}, (16) \cite{2010ChenL1448}, (17) \cite{2018MarcelinoALMA}, (18) \cite{2020FujishiroLow}, (19) \cite{2007HatchellStar}

\raggedright $^{a}$ Maximum projected values. \\
\raggedright $^{b}$ Projected distance between the continuum peak and the closest peak of the integrated intensity emission of a deuterated specie. The deuterated specie(s) and transition is in parenthesis. If more than one, the average is reported. \\
\raggedright $^{c}$ Separation is within the beam size of the observation. \\
\raggedright $^{d}$ \cite{2020FujishiroLow} reported no significant difference between the peaks of the N$_2$D$^{+}$(3-2) integrated intensity and the 1.3 mm continuum. The value we report here for the separation corresponds to one third of the minor axis of their observations. 

\end{table*}

\subsection{Herschel PACS/SPIRE fluxes for Per-bolo 45, Per-bolo 58 and L1451-mm}

We performed photometry on the Herschel 160, 250, 350, and 500 $\mu$m observations (\citealt{2012PezzutoHerschel,2012SadavoyHerschel}) for Per-bolo 45, Per-bolo 58 and L1451-mm as we could not find these measurements in the literature. Observations were taken using the PACS/SPIRE parallel fast mode, for which the typical photometric uncertainties are given in Table 1 from \cite{2015GuzmanFar}. The semi-major axes of the apertures at 250 $\mu$m were taken as 25"$\times$25", 20"$\times$25", and 25"$\times$25" for Per-bolo 45, Per-bolo 58 and L1451-mm, respectively. The P.A for Per-bolo 58 elliptical aperture was 25$^{\circ}$. The width for the background annulus was taken as 9" for all sources. The apertures at 160, 350 and 500 $\mu$m were calculated based on the apertures values at 250 $\mu$m. If the aperture major and minor axes at 250$\mu$m are A and B, then the apertures axes at the other wavelengths were calculated as $\sqrt{A^2-\theta^2_{250}+\theta^2_{wv}}$ and $\sqrt{B^2-\theta^2_{250}+\theta^2_{wv}}$ where $\theta_{wv}$ is the FWHM of the Herschel beam at a wavelength $wv$. The latter correspond to 12", 17", 24" and 35" for the wavelengths 160, 250, 350 and 500 $\mu$m, respectively. Errors were calculated as the sum of the 10\% calibration error and the 1$\sigma$ point source sensitivity taken as 12 mJy for SPIRE measurements and 33 mJy for PACS measurements\footnote{\url{http://herschel.esac.esa.int/Docs/PMODE/html/ch02s03.html}}. The assumed 10\% calibration error is a conservative approximation of the combined uncertainty in the flux calibration for point sources (5.5 \% and 7\% for SPIRE and PACS bands, respectively) and an additional ($\sim$ 4 \%) uncertainty for extended emission due to uncertainty in the measured beam area \citep{2013BendoFlux,2013GriffinFlux,2014BalogHerschel}. Table~\ref{tb:phot} summarizes the measured fluxes at these wavelengths for Per-bolo 45, Per-bolo 58 and L1451-mm. In addition, only Per-bolo 58 was detected in the PACS band at 70 $\mu$m, for which the 1$\sigma$ point source sensitivity is about 20 mJy. The measured flux for Per-bolo 58 at 70 $\mu$m was ~0.1 Jy, in agreement with the results of \cite{2010EnochCandidate} based on Spitzer observations ($\sim$0.065 Jy).

\begin{table*}

\caption{Herschel photometry}
\label{tb:phot}
\begin{tabular}{lccc}
\hline
\hline
Source & wavelength ($\mu$m) &Flux (Jy) & Flux error (Jy)\\

\hline
\hline
Per-bolo 45 &160 & 1.1 & 0.1\\
&250 & 9.5 & 1.0\\
&350 & 6.7 &0.7\\
&500 & 6.4 & 0.6\\
Per-bolo 58 &160 & 1.8 & 0.2\\
&250& 2.9 & 0.3\\
&350& 3.2 & 0.3\\
&500& 1.7 & 0.2\\
L1451-mm &160 & 0.4 & 0.1 \\
&250 & 1.5 & 0.2\\
&350 & 2.8 & 0.3\\
&500 & 1.7 & 0.2\\
\hline
\hline
\end{tabular}
\end{table*}


\bsp	
\label{lastpage}
\end{document}